\documentclass[11pt]{article}
\usepackage{amssymb,amsfonts,amsmath,latexsym}
\usepackage{color,float,fancyhdr}
\usepackage{graphicx}
\usepackage{wrapfig}
\usepackage{lscape}
\usepackage{rotating}
\usepackage{booktabs}
\usepackage[table]{xcolor}

\graphicspath{{Figures/}}
\usepackage[latin1]{inputenc}
\usepackage{pstricks,pst-node,pst-text,pst-3d}

\usepackage{natbib}

\usepackage{ifpdf}

\usepackage{color}
\definecolor{myred}{rgb}{0.5,0,0}
\definecolor{myblue}{rgb}{0,0,0.75}
\definecolor{mygreen}{rgb}{0,0.5,0}

\newcommand{\E}{\mathbb{E}}

\newcommand{\VaR}{\operatorname{VaR}}
\newcommand{\ES}{\operatorname{ES}}
\newcommand{\var}{\operatorname{var}}

\usepackage{vmargin}

\usepackage{graphicx}
\parskip\baselineskip
\usepackage[english]{babel}
\usepackage{blindtext}
\usepackage{sectsty}
\usepackage[scaled=.60]{beramono}
\usepackage{courier}
\usepackage{parskip}
\usepackage{etoolbox}
\newcounter{subeqn} %

\usepackage[font={small,it}]{caption}

\def\E{{\rm E}\,}
\def\P{{\rm P}\,}

\def\a{\alpha}

\def\g{\gamma}

\sectionfont{\fontsize{12}{15}\selectfont}
\subsectionfont{\fontsize{12}{15}\selectfont}

\numberwithin{equation}{section}

\title{Multinomial VaR Backtests: \\ A simple implicit approach to backtesting expected shortfall}

\author{Marie Kratz\thanks{ESSEC Business School, CREAR risk research center; \, E-mail: kratz@essec.edu }, Yen H. Lok \thanks{Heriot Watt University; \,  E-mail: yhl30@hw.ac.uk}, Alexander J.~McNeil \thanks{University of York; \,  E-mail: alexander.mcneil@york.ac.uk} }

\date{}

\begin{document}

\maketitle

\begin{abstract}
\noindent Under the Fundamental Review of the Trading Book (FRTB) capital charges for the trading book are based on the coherent expected shortfall (ES) risk measure, which show greater sensitivity to tail risk. In this paper it is argued that backtesting of expected shortfall - or the trading book model from which it is calculated - can be based on a simultaneous multinomial test of value-at-risk (VaR) exceptions at different levels, an idea supported by an approximation of ES in terms of multiple quantiles of a distribution proposed in~\citet{bib:emmer-kratz-tasche-15}. By comparing Pearson, Nass and likelihood-ratio tests (LRTs) for different numbers of VaR levels $N$ it is shown in a series of simulation experiments that multinomial tests with $N\geq 4$ are much more powerful at detecting misspecifications of trading book loss models than standard binomial exception tests corresponding to the case $N=1$. Each test has its merits: Pearson offers simplicity; Nass is robust in its size properties to the choice of $N$; the LRT is very powerful though slightly over-sized in small samples and more computationally burdensome. A traffic-light system for trading book models based on the multinomial test is proposed and the recommended procedure is applied to a real-data example spanning the 2008 financial crisis.
\vspace{0.7ex}\\
\emph{2010 AMS classification}: 60G70; 62C05; 62P05; 91B30; 91G70; 91G99  
\end{abstract}

\section{Introduction}\label{se:intro}

Techniques for the measurement of risk are central to the process of managing risk in financial institutions and beyond. In banking and insurance it is standard to model risk with probability distributions and express risk in terms of scalar-valued risk measures.
Formally speaking, \emph{risk measures} are mappings of random variables representing profits and losses (P\&L) into real numbers representing capital amounts required as a buffer against insolvency. 

There is a very large literature on risk measures and their properties, and we limit our survey to key references that have had an impact on practice and the regulatory debate. In a seminal work \citet{bib:artzner-et-al-99} proposed a set of desirable mathematical properties defining a \emph{coherent} risk measure, important axioms being \emph{subadditivity}, which is essential to measure the diversification benefits in a risky portfolio, and \emph{positive homogeneity}, which requires a linear scaling of the risk measure with portfolio size. \citet{bib:foellmer-schied-02b} defined the larger class of \emph{convex} risk measures by replacing the subadditivity and positive homogeneity axioms by the weaker requirement of convexity; see also \citet{bib:foellmer-schied-11}. 
%


The two main risk measures used in financial institutions and regulation are value-at-risk (VaR) and expected shortfall (ES), the latter also known as tail value-at-risk (TVaR). VaR is defined as a quantile of the P\&L distribution and, despite the fact that it is neither coherent nor convex, it has been the dominant risk measure in banking regulation. It is also the risk measure used in Solvency II insurance regulation in Europe, where the Solvency Capital Requirement (SCR) is defined to be the 99.5\% VaR of an annual loss distribution.

Expected shortfall at level $\alpha$ is the conditional expected loss given exceedance of VaR at that level and is a coherent risk measure \citep{bib:acerbi-tasche-02,bib:tasche-02}. For this reason, and also because it is a more tail-sensitive measure of risk, it has attracted increasing regulatory attention in recent years. ES at the 99\% level for annual losses is the primary risk measure in the Swiss Solvency Test (SST). As a result of the Fundamental Review of the Trading Book~\citep{bib:basel-13} a 10-day ES at the 97.5\% level will be the main measure of risk for setting trading book capital under Basel III~\citep{bib:basel-16}.

For a given risk measure, it is vital to be able to estimate it accurately, and to validate estimates by checking whether realized losses, observed ex post, are in line with the ex ante estimates or forecasts. The statistical procedure by which we compare realizations with forecasts is known as \emph{backtesting}. 

The literature on backtesting VaR estimates is large and is based on the observation that when VaR at level $\alpha$ is consistently well estimated the \emph{VaR exceptions}, that is the occasions on which realized losses exceed VaR forecasts, should form a sequence of independent, identically distributed (iid) Bernoulli variables with probability $(1-\alpha)$. 

An early paper by \citet{bib:kupiec-95} proposed
a binomial likelihood ratio test for the number of exceptions as well a test
based on the fact that the spacings between violations should be
geometrically distributed; see also \citet{bib:dave-stahl-98}. The simple binomial test for the number of violations is often described
as a test of \emph{unconditional} coverage, while a test that also explicitly
examines the independence of violations is a test of \emph{conditional} coverage.
 \citet{bib:christoffersen-98} proposed a test of conditional coverage
 in which the iid Bernoulli hypothesis is tested against the
 alternative hypothesis that violations show dependence characterized by first-order Markov
behaviour; see also the recent paper by
\citet{bib:davis-13}. \citet{bib:christoffersen-pelletier-04} further developed the idea of testing
the spacings between VaR violations using the fact that a discrete
geometric distribution can be
approximated by a continuous exponential distribution. The null
hypothesis of exponential spacings (constant hazard model) is tested
against a Weibull alternative (in which the hazard function may be
increasing or decreasing). \citet{bib:berkowitz-christoffersen-pelletier-11} provide a
comprehensive overview of tests of conditional coverage. They
advocate, in particular, the geometric test and a 
regression test based on an idea developed by \citet{bib:engle-manganelli-04}
for checking the fit of the CAViaR model for dynamic quantiles.

The literature on ES backtesting is smaller. \cite{bib:mcneil-frey-00}  suggest a bootstrap
hypothesis test based on so-called violation
residuals. These measure the discrepancy between the realized losses and
the expected shortfall estimates on days when VaR violations
take place and should form a sample from a
distribution with mean zero.
\cite{bib:acerbi-szekely-14} look at similar kinds of statistics and suggest the use of
Monte Carlo hypothesis tests. Recently \citet{bib:costanzino-curran-15} have proposed a Z-test for a discretized version of expected shortfall which extends to other so-called spectral risk measures; see also \citet{bib:costanzino-curran-16} where the idea is extended to propose a traffic light system analogous to the Basel system for VaR exceptions.


A further strand of the backtesting literature looks at backtesting methods based on \emph{realized $p$ values} or probability-integral-transform (PIT) values. These are estimates of the probability of observing a particular ex post loss based on the predictive density from which risk measure estimates are derived; they should form an iid uniform sample when ex ante models are consistent with ex post losses.
Rather than focussing on point estimates of risk measures,~\citet{bib:diebold-gunther-tay-98} show how realized $p$-values can be used to evaluate the overall quality of density forecasts. In \citet{bib:diebold-hahn-tay-99}, the authors extended the density forecast evaluation to the multivariate case. \citet{bib:blum-04} studied various issues left open, and proposed and validated mathematically a method based on PIT also in situations with overlapping forecast intervals and multiple forecast horizons. \citet{bib:berkowitz-01} proposed a test of the quality of the tail of the predictive model based on the idea of truncating realized $p$-values above a level $\alpha$. A backtesting
procedure for expected shortfall based on realized $p$-values may be found in
\citet{bib:kerkhof-melenberg-04}. 


Some authors have cast doubt on the feasibility of backtesting expected shortfall. It has been shown that estimators of ES generally lack robustness \citep{bib:cont-deguest-scandolo-10} so stable series of ES estimates are more difficult to obtain than VaR estimates. However,~\citet{bib:emmer-kratz-tasche-15} point our that the concept of robustness, which was introduced in statistics in the context of measurement error, may be less relevant in finance and insurance, where extreme values often occur as part of the data-generating process and not as outliers or measurement errors; they argue that (particularly in insurance) large outcomes may actually be more accurately monitored than smaller ones, and their values better estimated.

\citet{bib:gneiting-11} showed that ES is not an \emph{elicitable} risk measure, whereas VaR is; see also \citet{bib:bellini-bignozzi-13} and \citet{bib:ziegel-13} on this subject. An elicitable risk measure is a statistic of a P\&L distribution that can be represented as the solution of a forecasting-error minimization problem. The concept was introduced by \citet{bib:osband-85} and \citet{bib:lambert-pennock-shoham-08}. When a risk measure is elicitable we can use \emph{consistent} scoring functions to compare series of forecasts obtained by different modelling approaches and obtain objective guidance on the approach that gives the best forecasting performance.

Although \citet{bib:gneiting-11} suggested that the lack of elicitability of ES called into question our  ability to backtest ES and its use in risk management, a consensus is now emerging that the problem of \emph{comparing} the forecasting performance of different estimation methods is distinct from the problem of addressing whether a single series of ex ante ES estimates is consistent with a series of ex post realizations of P\&L, and that there are reasonable approaches to the latter problem as mentioned above. There is a large econometrics literature on comparitive forecasting performance inlcuding~\citet{bib:diebold-mariano-95} and~\citet{bib:giacomini-white-06}.

It should be noted that ES satisfies more general notions of elicitability, such as \emph{conditional elicitability} and \emph{joint elicitability}. \citet{bib:emmer-kratz-tasche-15} introduced the concept of conditional elicitability. This offers a way of splitting a forecasting method into two component methods involving elicitable statistics and separately backtesting and comparing their forecast performances. Since ES is the expected loss conditional on exceedance of VaR, we can first backtest VaR using an appropriate consistent scoring function and then, treating VaR as a fixed constant, we can backtest ES using the squared error scoring function for an elicitable mean. \citet{bib:fissler-ziegel-15} show that VaR and ES are jointly elicitable in the sense that they jointly minimize an appropriate bi-dimensional scoring function; this allows the comparison of different forecasting methods that give estimates of both VaR and ES. See also~\citet{bib:acerbi-szekely-16} who introduce a new concept of ``backtestability'' satisfied in particular by expected shortfall.
%


In this paper our goal is to propose a simple approach to backtesting which may be viewed in two ways: on the one hand as a natural extension to standard VaR backtesting that allows us to test VaR estimates at different $\alpha$ levels \emph{simultaneously} using a multinomial distribution; on the other hand as an \emph{implicit} backtest for ES. 

Although the FRTB has recommended that ES be adopted as the main risk measure for the trading book under Basel III \citep{bib:basel-16}, it is notable that the backtesting regime will still largely be based on looking at VaR exceptions at the 99\% level, albeit also for individual trading desks as well as the whole trading book. The Basel publication does however state that banks will be required to go beyond the basic mandatory requirement to also consider more advanced backtests. They list a number of possibilities including: tests based on VaR at multiple levels (they explicitly mention 97.5\% and 99\%); tests based on both VaR and expected shortfall; tests based on realized $p$-values.

The idea that our test serves as an implicit backtest of expected shortfall comes naturally from an approximation of ES proposed by \cite{bib:emmer-kratz-tasche-15}. Denoting the ES and VaR of the distribution of the loss $L$ by $\ES_\alpha(L)$ and $\VaR_\alpha(L)$, these authors suggest the following approximation of ES:
\begin{equation}\label{eq:approxEKT15}
\mathrm{ES}_\alpha(L) \; \approx \; \frac14 \,\left[\, q(\alpha)+q(0.75\,\alpha+0.25) +  q(0.5\,\alpha+0.5)+q(0.25\,\alpha+0.75)\,\right]
\end{equation}
where $q(\gamma)= VaR_\gamma(L)$.
This suggests that an estimate of $\mathrm{ES}_\alpha(L)$ derived from a model for the distribution of $L$ could be considered reliable if estimates of the four VaR values $q(a\alpha+b)$ derived from the same model are reliable. It leads to the intuitive idea of backtesting ES via simultaneously backtesting multiple VaR estimates at different levels.

In this paper we propose multinomial tests of VaR exceptions at multiple levels, examining the properties of the tests and answering the following main  questions: 
\begin{itemize}
\item Does a multinomial test work better than a binomial one for model validation? 
\item Which particular form of the multinomial test should we use in which situation? 
\item  What is the optimal number of quantiles that should be used in terms of size, power and stability of results, as well as  simplicity of the procedure? 
\end{itemize}

A guiding principle of our study is to provide a \emph{simple} test that is not much more complicated (conceptually and computationally) than the binomial test based on VaR exception counts, which dominates industry and regulatory practice. Our test should be more powerful than the binomial test and better able to reject models that give poor estimates of the tail, and which would thus lead to poor estimates of expected shortfall. However, maximizing power is not the overriding concern. Our proposed backtest may not necessarily attain the power of other tests based on realized $p$-values, but it gives impressive results nontheless and we believe it is a much easier test to interpret for practitioners and regulators. It also leads to a very intuitive traffic-light systems for model validation that extends and improves the existing Basel traffic-light system.


The structure of the paper is as follows. The multinomial backtest is defined in Section~\ref{sec:mtest} and three variants are proposed: the standard Pearson chi-squared test; the Nass test; a likelihood ratio test (LRT). We also show how the latter relates to a test of~\citet{bib:berkowitz-01} based on realized $p$-values. A large simulation study in several parts is presented in Section~\ref{sec:numerical}. This contains a study of the size and power of the multinomial tests, where we look in particular at the ability of the tests to discriminate against models that underestimate the kurtosis and skewness of the loss distribution. We also conduct static (distribution-based) and dynamic (time-series based) backtests in which we show how fictitious forecasters who estimate models of greater and lesser quality would be treated by the multinomial tests.

Based on the results of Section~\ref{sec:numerical}, we give our views on the best design of a simultaneous backtest of VaR at several levels, or equivalently an implicit backtest of expected shortfall, in Section~\ref{sec:proc-impl-backt}. We show also how a traffic-light system may be designed. In Section~\ref{sec:empirical}, we apply the method to real data, considering the Standard \& Poor's 500 index. Conclusions are found in Section~\ref{sec:conclusion}.

\section{Multinomial tests} \label{sec:mtest}

\subsection{Testing set-up}
Suppose we have a series of ex-ante predictive models $\{F_t, t = 1, \ldots, n\}$ and a series of ex-post losses $\{L_t, t = 1, \ldots, n\}$. At each time $t$ the model $F_t$ is used to produce estimates (or forecasts) of value-at-risk $\VaR_{\alpha,t}$ and expected shortfall $\ES_{\alpha,t}$ at various probability levels $\alpha$. The VaR estimates are 
compared with $L_t$ to assess the adequacy of the models in describing the losses, with particular emphasis on the most extreme losses.

In view of the representation~\eqref{eq:approxEKT15}, we consider the idea proposed in~\citet{bib:emmer-kratz-tasche-15} of backtesting the ES estimates indirectly by simultaneously backtesting a number of VaR estimates at different levels $\alpha_1,\ldots,\alpha_N$. We investigate different choices of the number of levels $N$ in the simulation study in Section~\ref{sec:numerical}. 

We generalize the idea of~\eqref{eq:approxEKT15} by considering VaR probability levels $\alpha_1,\ldots,\alpha_N$ defined by
\begin{equation}\label{eq: approxES}
\a_j = \a + \frac{j-1}{N}(1-\alpha), \quad j=1,\ldots,N,\;\;N \in \mathbb{N},
\end{equation}
for some starting level $\alpha$. In this paper we generally set $\alpha=0.975$ corresponding to the level used for expected shortfall calculation and the lower of the two levels used for backtesting under the Basel rules for banks~\citep{bib:basel-16}; we will also consider $\alpha=0.99$ in the case when $N=1$ since this is the usual level for binomial tests of VaR exceptions. To complete the description of levels we set $\alpha_0=0$ and $\alpha_{N+1}=1$.

We define the violation or exception indicator of the level $\alpha_j$ at time $t$ by 
\begin{equation}\label{eq:violationPcs}
	I_{t,j} :=I_{\{L_t> \VaR_{\a_j,t}\}}
\end{equation}
where $I_A$ denotes an event indicator for the event $A$.

It is well known~\citep{bib:christoffersen-98} that if the losses $L_t$ have conditional distribution functions $F_t$ then, for fixed $j$, the sequence $(I_{t,j})_{t=1,\ldots,n}$ should satisfy:
\begin{itemize}
	\item the \emph{unconditional coverage} hypothesis, $\E(I_{t,j})=1-\alpha_j$ for all $t$, and
	\item the \emph{independence hypothesis}, $I_{t,j}$ is independent of $I_{s,j}$ for $s\ne t$.
\end{itemize}
If both are satisfied the VaR forecasts at level $\alpha_j$ are said to satisfy the hypothesis of correct \emph{conditional coverage} and the number of exceptions $\sum_{t=1}^n I_{t,j}$ has a binomial distribution with success (exception) probability $1-\a_j$.

Testing simultaneously VaR estimates at $N$ levels leads to a multinomial distribution. If we define $X_t = \sum_{j=1}^N I_{t,j}$ then the sequence $(X_t)_{t=1,\ldots,n}$ counts the number of VaR levels that are breached. The sequence $(X_t)$ should satisfy the two conditions: 
\begin{itemize}
	\item the \emph{unconditional coverage} hypothesis, $\P(X_t \leq j) = \alpha_{j+1},\;j=0,\ldots,N$ for all $t$, 
	\item the \emph{independence hypothesis}, $X_{t}$ is independent of $X_{s}$ for $s\ne t$.
\end{itemize}
The unconditional coverage property can also be written 
$$
X_t\sim \text{MN}(1,(\alpha_1-\alpha_0,\ldots,\alpha_{N+1}-\alpha_N)), \; \text{for all}\; t.
$$
Here $\text{MN}(n,(p_0,\ldots,p_N))$ denotes the multinomial distribution with $n$ trials, each of which may result in one of $N+1$ outcomes $\{0,1,\ldots,N\}$ according to probabilities $p_0,\ldots,p_N$ that sum to one. 
If we now define observed cell counts by
\begin{displaymath}
O_j = \sum_{t=1}^n I_{\{X_t=j\}},\quad j=0,1\ldots,N,
\end{displaymath}
then the random vector $\displaystyle (O_0,\ldots,O_{N})$ should follow the multinomial distribution
\begin{displaymath}
  (O_0,\ldots,O_{N}) \sim \text{MN}(n,(\alpha_1-\alpha_0,\ldots,\alpha_{N+1}-\alpha_N))\;\;.
\end{displaymath}
More formally, let $0=\theta_0<\theta_1< \cdots < \theta_N <\theta_{N+1}=1$ be an arbitrary sequence of parameters and consider the model where $(O_0,\ldots,O_{N}) \sim \text{MN}(n,(\theta_1-\theta_0,\ldots,\theta_{N+1}-\theta_N))$. We test null and alternative hypotheses given by
\begin{equation}\label{eq:H0bis}
\left|\begin{array}{lccl}
 \text{H}0: & \theta_j &=\alpha_j & \text{for} \;j=1,\ldots,N \\
\text{H}1: &\theta_j &\neq \alpha_j &\text{for at least one}\; j\in\{1,\ldots,N\}.
\end{array}\right.
\end{equation}

\subsection{Choice of tests}

Various test statistics can be used to evaluate these hypotheses. \citet{bib:cai-krishnamoorthy-06} provide a relevant numerical study of the properties of five possible tests of multinomial proportions. Here we propose to use three of them: the standard Pearson chi-square test; the Nass test, which performs better with small cell counts; a likelihood ratio test (LRT). More details are as follows.
\begin{enumerate}
\item {\it Pearson chi-squared test}~\citep{bib:pearson-00}.
The test statistic in this case is
\vspace{-1ex}
\begin{equation}\label{def:S_N}
S_N= \sum_{j=0}^N \frac{(O_{j+1} - n(\a_{j+1} - \a_{j}))^2}{n(\a_{j+1} - \a_{j})} \, \underset{H0}{\overset{d}{\thicksim}}\,  \chi^2_{N}
\end{equation}
and a size $\kappa$ test is obtained by rejecting the null hypothesis when $S_N > \chi^2_{N}  (1-\kappa)$,
where $\chi^2_{N}(1-\kappa)$ is the $(1-\kappa)$-quantile of the $\chi^2_{N}$-distribution. It is well known that the accuracy of this test increases as $\displaystyle \min_{0 \le j\le N} n(\a_{j+1} - \a_{j})$ increases and decreases with increasing $N$.

\item {\it Nass test}~\citep{bib:nass-59}.

Nass introduced an improved approximation to the distribution of the statistic $S_N$ defined in \eqref{def:S_N}, namely 
$$
c\, S_N \; \underset{H0}{\overset{d}{\thicksim}}\;  \chi^2_\nu \,,\quad\text{with}\quad c= \frac{2\,\E(S_N)}{\var(S_N)} \; \,\text{and}\;\, \nu= c\,\E(S_N),
$$
where $\displaystyle \E(S_N)=N$ and 
$\displaystyle \var(S_N)= 2N-\frac{N^2+4N+1}{n}+\frac1n \sum_{j=0}^N \frac1{\a_{j+1} - \a_{j} }$ .

The null hypothesis is rejected when $\displaystyle c\,S_N > \chi^2_\nu  (1-\kappa)$,  using the same notation as before. The Nass test offers an appreciable improvement over the chi-square test when cell probabilities are small.

\item {\it LRT} (see, for example, \citet{bib:casella-berger-02}).

In a LRT we calculate maximum likelihood estimates $\hat{\theta}_j$ of the parameters $\theta_j$ under the alternative hypothesis H1 and we form the statistic
\begin{displaymath}
 \tilde{S}_N= 2 \sum_{j=0}^{N} O_{j} \ln\left(\frac{\hat{\theta}_{j+1}-\hat{\theta}_{j}}{\alpha_{j+1}-\alpha_{j}}\right).
\end{displaymath}
 Under the unrestricted multinomial model $(O_0,\ldots,O_{N}) \sim \text{MN}(n,(\theta_1-\theta_0,\ldots,\theta_{N+1}-\theta_N))$ the estimated multinomial cell probabilities are given by $\hat{\theta}_{j+1}-\hat{\theta}_{j} = O_j/n$, and are thus zero when $O_j$ is zero, which leads to an undefined test statistic.

For this reason, whenever $N\geq 2$, we use a different version of the LRT to the one described in~\cite{bib:cai-krishnamoorthy-06}. We consider a general model in which the parameters are given by
\begin{equation}\label{eq:4}
  \theta_j = \Phi\left(\frac{\Phi^{-1}(\alpha_j)-\mu}{\sigma}\right), \quad j=1,\ldots,N,
\end{equation}
where $\mu\in \mathbb{R}$, $\sigma >0$ and $\Phi$ denotes the standard normal distribution function. In the restricted model we test the null hypothesis H0: $\mu=0$ and $\sigma=1$ against the alternative H1: $\mu\neq 0$ or $\sigma\neq 1$.  In this case we have
\begin{displaymath}
 \hat{\theta}_{j+1}-\hat{\theta}_{j} =
\Phi\left(\frac{\Phi^{-1}(\alpha_{j+1})-\hat{\mu}}{\hat{\sigma}}\right)- \Phi\left(\frac{\Phi^{-1}(\alpha_j)-\hat{\mu}}{\hat{\sigma}}\right),
\end{displaymath}
where $\hat{\mu}$ and
$\hat{\sigma}$ are the MLEs under H1, so that the problem of zero estimated cell probabilities does not arise. The test statistic $G_N$ is asymptotically chi-squared distributed with two degrees of freedom 
and the null is rejected if $G_N > \chi^2_2 (1-\kappa)$. 

\end{enumerate}

\subsection{The case $N=1$}\label{sec:case-n=1}
In the case where $N=1$ we carry out an augmented set of binomial tests. For the LRT in the case $N=1$, there is only one free parameter to determine ($\theta_1$) and we carry out a standard two-sided asymptotic likelihood ratio test against the unrestricted alternative model; in this case the statistic is compared to a $\chi^2_1$-distribution.

It may be easily verified that, for $N=1$, the Pearson multinomial test statistic $S_1$ in~\eqref{def:S_N} is the square of the binomial \emph{score} test statistic
 \begin{equation}\label{eq:3}
 Z :=   \frac{n^{-1}\sum_{t=1}^n
  I_{t,1} - (1-\alpha)}{\sqrt{n^{-1}\alpha(1-\alpha)}} =   \frac{O_1
   - n(1-\alpha)}{\sqrt{n\alpha(1-\alpha)}},
  \end{equation}
which is compared with a standard normal distribution; thus a two-sided score test will give identical results to the Pearson chi-squared test in this case. In addition to the score test we also consider a \emph{Wald} test in which the $\alpha$ parameter in the denominator of~\eqref{eq:3} is replaced by the estimator $\hat{\theta}_1 = n^{-1}\sum_{t=1}^n(1- I_{t,1}) = 1- O_1/n$.

As well as two-sided tests, we carry out one-sided variants of the LRT, score and Wald tests which test $H0 :  \theta_1 \geq \alpha$ against the alternative $H1 : \theta_1 < \alpha$ (underestimation of VaR). One-sided score and Wald tests are straightforward to carry out, being based on the asymptotic normality of $Z$. To derive a one-sided LRT it may be noted that the likelihood ratio statistic for testing the simple null hypothesis $\theta_1=\alpha$ against the simple alternative that $\theta_1 = \alpha^*$ with $\alpha^* < \alpha$ depends on the data through the the number of VaR exceptions $B=\sum_{t=1}^n I_{t,1}$. In the one-sided LRT we test $B$ against a binomial distribution; this test at the 99\% level is the one that underlies the Basel backtesting regime and traffic light system.

\subsection{The limiting multinomial LRT}
The multinomial LRT has a natural continuous limit as the number of levels $N$ goes to infinity, which coincides with a test proposed by~\citet{bib:berkowitz-01} based on \emph{realized $p$-values}.
Our LRT uses a multinomial model for $X_t^{(N)} := X_t = \sum_{j=1}^N I_{\{L_t > \VaR_{\alpha_j,t}\}}$ in which we assume that
\begin{equation}\label{eq:1}
  \P\left(X_t^{(N)} \leq j\right) = \theta_{j+1} = \Phi\left(\frac{\Phi^{-1}(\alpha+\tfrac{j}{N}(1-\alpha))-\mu}{\sigma}\right), j=0,\ldots,N,
\end{equation}
and in which we test for $\mu=0$ and $\sigma=1$.

The natural limiting model as $N\to\infty$ is based on the random variable $\displaystyle W_t^\alpha = (1-\alpha)^{-1}\int_\alpha^1 I_{\{L_t > \VaR_{u,t}\}}du$. For simplicity let us assume that $F_t$ is a continuous and strictly increasing distribution function and that $\VaR_{u,t} = F_t^{-1}(u)$ so that the event $\{L_t > \VaR_{u,t}\}$ is identical to the event $\{U_t > u\}$ where $U_t = F_t(L_t)$ is known as a realized $p$-value or a PIT (probability integral transform) value. If the losses $L_t$ have conditional distribution functions $F_t$ then the $U_t$ values should be iid uniform by the transformation of~\cite{bib:rosenblatt-52}. It is easily verified that
\begin{displaymath}
  W_t^\alpha = \int_\alpha^1 I_{\{L_t > \VaR_{u,t}\}}du = \int_\alpha^1 I_{\{U_t > u\}} du = \frac{\max(U_t,\alpha)-\alpha}{1-\alpha} \,.
\end{displaymath}
\citet{bib:berkowitz-01} proposed a test in which $Z_t^\alpha = \Phi^{-1}(\max(U_t,\alpha))$ is modelled by a truncated normal, that is a model where
\begin{equation}\label{eq:2}
  \P(Z_t^\alpha \leq z) = \Phi\left(\frac{z-\mu}{\sigma}\right), \quad z \geq \Phi^{-1}(\alpha),
\end{equation}
and in which we test for $\mu=0$ and $\sigma=1$ to assess the uniformity of the realized $p$-values with emphasis on the tail (that is above $\alpha$).
Since $W_t^\alpha = (\Phi(Z_t^\alpha)-\alpha)/(1-\alpha)$, the Berkowitz model~\eqref{eq:2} is equivalent to a model where
\begin{displaymath}
  \P(W_t^\alpha \leq w) = \Phi\left(\frac{\Phi^{-1}(\alpha+w(1-\alpha))-\mu}{\sigma}\right), \quad w \in [0,1),
\end{displaymath}
which is the natural continuous counterpart of the discrete model in~\eqref{eq:1}.

\section{Simulation studies}\label{sec:numerical}

We recall that the main questions of interest are: Does a multinomial test work better than a binomial one for model validation in terms of its size and power properties? Which of the three multinomial tests should be favoured in which situations? What is the optimal number of quantiles that should be used to obtain a good performance? 

To answer these questions, we conduct a series of experiments based on simulated data. In Section~\ref{sec:size-power} we carry out a  comparison of the size and power of our tests. The power experiments consider misspecifications of the loss distribution using distributional forms that might be typical for the trading book; we are particularly interested to see whether the multinomial tests can distinguish more effectively than binomial tests between distributions with different tails.

In Sections~\ref{sec:stat-backt-exper} and~\ref{sec:dyn} we carry out \emph{backtesting} experiments in which we look at the ability of the tests to distinguish between the performance of different modellers who estimate quantiles with different methodologies and are subject to statistical error. The backtests of Section~\ref{sec:stat-backt-exper} take a static distributional view; in other words the true data generating process is simply a distribution as in the size-power comparisons of Section~\ref{sec:size-power}.

In Section~\ref{sec:dyn} we take a dynamic view and consider a data-generating process which features a GARCH model of stochastic volatility with heavy-tailed innovations. We consider the ability of the multinomial tests to distinguish between good and bad forecasters, where the latter may misspecify the form of the dynamics and/or the conditional distribution of the losses.

\subsection{Size and Power}\label{sec:size-power}

\subsubsection{Theory}
To judge the effectiveness of the three multinomial tests (and the additional binomial tests), we compute their size $\g=\P(\text{reject H0} | \text{H0 true})$ (type I error) and power $1-\beta=1-\P(\text{accept H0} | \text{H0 false}) $ (1- type II error). For a given size, regulators should clearly be interested in having the most powerful tests for exposing banks working with deficient models.

Checking the size of the multinomial test requires us to simulate data
from a multinomial distribution under the null hypothesis (H0). This can be done indirectly by simulating data from any continuous distribution (such as normal) and counting the observations between the true
values of the $\alpha_j$-quantiles.

To calculate the power, we have to simulate data from multinomial models under the alternative hypothesis (H1). We choose to simulate from models where the parameters are given by
\begin{displaymath}
  \theta_j = G\left(F^{\leftarrow}(\alpha_j)\right)
\end{displaymath}
where $F$ and $G$ are distribution functions, $F^\leftarrow(u) = \inf\{x :F(x) \geq u\}$ denotes the generalized inverse of $F$, and $F$ and $G$ are chosen such that $\theta_j \neq \alpha_j$ for one or more values of $j$. $G$ can be thought of as the true distribution and $F$ as the model. If a forecaster uses $F$ to determine the $\alpha_j$-quantile, then the true probability associated with the quantile estimate is $\theta_j$ rather than $\alpha_j$. We consider the case where $F$ and $G$ are two different distributions with mean zero and variance one, but different shapes. 

In a time-series context we could think of the following situation. Suppose that the losses $(L_t)$ form a time series adapted to a filtration $(\mathcal{F}_t)$ and that, for all $t$, the true conditional distribution of $L_t$ given $\mathcal{F}_{t-1}$ is given by $G_t(x) = G((x-\mu_t)/\sigma_t)$ where $\mu_t$ and $\sigma_t$ are $\mathcal{F}_{t-1}$-measurable variables representing the conditional mean and standard deviation of $L_t$. 
However a modeller uses the model $F_t(x) =F((x-\mu_t)/\sigma_t)$ in which the distributional form is misspecified but the conditional mean and standard deviation are correct. He thus delivers VaR estimates given by $\VaR_{\alpha,t}= \mu_t+\sigma_tF^{\leftarrow}(\alpha_j)$. The true probabilities associated with these VaR estimates are 
$\theta_j = G_t(\VaR_{\alpha_j,t}) =G(F^{\leftarrow}(\alpha_j))\neq \alpha_j$. We are interested in discovering whether the tests have the power to detect that the forecaster has used the models $\{F_t,t=1,\ldots,n\}$ rather than the true distributions $\{G_t,t=1,\ldots,n\}$.

Suppose for instance that $G$ is a Student t distribution (scaled to have unit variance) and $F$ is a
normal so that the forecaster underestimates the more extreme
quantiles. In such a case, we will tend to observe too many exceedances of the higher quantiles.

The size calculation corresponds to the situation where $F$ = $G$; we calculate quantiles using the true model and there is no misspecification.
In the power calculation we focus on distributional forms for $G$ that are typical for the trading book, having heavy tails and possibly skewness. We consider Student distributions with 5 and 3 degrees of freedom ($t5$ and $t3$) which have moderately heavy and heavy tails respectively, and the skewed Student distribution of~\citet{bib:fernandez-steel-98} with 3 degrees of freedom and a skewness parameter $\gamma=1.2$ (denoted sk$t3$). In practice we simulate observations from $G$ and count the numbers lying between the $N$ quantiles of $F$; in all cases we take the benchmark model $F$ to be standard normal.

Table~\ref{table:rmvalues} shows the values of $\VaR_{0.975}$, $\VaR_{0.99}$ and $\ES_{0.975}$ for the four distributions used in the simulation study. These distributions have all been calibrated to have mean zero and variance one. 
Note how the value of $\ES_{0.975}$ get progressively larger as we move down the table; the final column marked $\Delta_2$ shows the percentage increase in the value of $\ES_{0.975}$ when compared with the normal distribution. Since capital is supposed to be based on this risk measure it is particularly important that a bank can estimate this measure reliably. From a regulatory perspective it is important that backtesting procedure can distinguish the heavier-tailed models from the light-tailed normal distribution since a modeller using the normal distribution would seriously underestimate $\ES_{0.975}$ if any of the other three distributions were ``the true distribution''. 

The three distributions give comparable values for $\VaR_{0.975}$; the t3 model actually gives the smallest value for this risk measure. The values of $\VaR_{0.99}$ are ordered in the same way as those of $\ES_{0.975}$. $\Delta_1$, which shows the percentage increase in the value of $\VaR_{0.99}$ when compared with the normal distribution, does not increase quite so dramatically as $\Delta_2$, which already suggests that more than two quantiles might be needed to implicitly backtest ES.

\begin{table}[htbp]
\centering
\begin{tabular}{rrrrrr}
  \toprule
 & $\VaR_{0.975}$ & $\VaR_{0.99}$ & $\Delta_{1}$ & $\ES_{0.975}$ & $\Delta_{2}$ \\ 
  \midrule
Normal & 1.96 & 2.33 & 0.00 & 2.34 & 0.00 \\ 
  t5 & 1.99 & 2.61 & 12.04 & 2.73 & 16.68 \\ 
  t3 & 1.84 & 2.62 & 12.69 & 2.91 & 24.46 \\ 
  st3 ($\gamma =1.2$) & 2.04 & 2.99 & 28.68 & 3.35 & 43.11 \\ 
   \bottomrule
\end{tabular}
\caption{Values of $\VaR_{0.975}$, $\VaR_{0.99}$ and $\ES_{0.975}$ for four distributions
used in simulation study (Normal, Student t5, Student t3, skewed Student t3
with skewness parameter $\gamma = 1.2$). $\Delta_1$ column shows percentage increase in
$\VaR_{0.99}$ compared with normal distribution; $\Delta_2$ column shows percentage increase in
$\ES_{0.975}$ compared with normal distribution.} 
\label{table:rmvalues}
\end{table}

To determine the VaR level values we set $N=2^k$ for $k=0, 1,\cdots, 6$. In all multinomial experiments with $N\geq 2$ we set $\alpha_1=\alpha=0.975$ and further levels are determined by~\eqref{eq: approxES}. We choose sample sizes $n_1=250, 500, 1000, 2000$ and estimate the rejection probability for the null hypothesis using 10'000 replications.

In the case $N=1$ we consider a series of additional binomial tests of the number of exceptions of the level $\alpha_1=\alpha$ and present these in a separate table; in this case we also consider the level $\alpha=0.99$ in addition to $\alpha=0.975$. This gives us the ability to compare our multinomial tests with all binomial test variants at both levels and thus to evaluate whether the multinomial tests are really superior to current practice. 

\begin{sidewaystable}[htbp]
  \centering
  \begin{tabular}{*{2}{l}*{12}{r}}
    \toprule
     & \( \alpha \) & \multicolumn{6}{c |}{0.975} & \multicolumn{6}{c}{0.990} \\
    \cmidrule(lr){3-8} \cmidrule(lr){9-14}
     & twosided & \multicolumn{3}{c|}{TRUE} & \multicolumn{3}{c|}{FALSE} & \multicolumn{3}{c|}{TRUE} & \multicolumn{3}{c}{FALSE} \\
    \cmidrule(lr){3-5} \cmidrule(lr){6-8} \cmidrule(lr){9-11} \cmidrule(lr){12-14}
    \( G \) & \( n \) \textbar\ test & \multicolumn{1}{c}{Wald} & \multicolumn{1}{c}{score} & \multicolumn{1}{c|}{LRT} & \multicolumn{1}{c}{Wald} & \multicolumn{1}{c}{score} & \multicolumn{1}{c|}{LRT} & \multicolumn{1}{c}{Wald} & \multicolumn{1}{c}{score} & \multicolumn{1}{c|}{LRT} & \multicolumn{1}{c}{Wald} & \multicolumn{1}{c}{score} & \multicolumn{1}{c}{LRT} \\
    \midrule
    Normal & 250 & \cellcolor{green} 5.7 & \cellcolor{green} 3.9 & 7.5 & \cellcolor{green} 2.4 & \cellcolor{green} 5.0 & \cellcolor{green} 5.0 & 8.0 & \cellcolor{green} 4.0 & 8.9 & \cellcolor{green} 1.2 & \cellcolor{green} 4.0 & \cellcolor{pink} 10.5 \\
    & 500 & 7.8 & \cellcolor{green} 3.9 & \cellcolor{green} 5.9 & \cellcolor{green} 2.6 & \cellcolor{green} 4.7 & 7.9 & \cellcolor{red} 12.5 & \cellcolor{green} 3.7 & 7.0 & \cellcolor{green} 1.3 & 6.7 & 6.7 \\
    & 1000 & \cellcolor{green} 5.0 & \cellcolor{green} 5.0 & \cellcolor{green} 4.1 & \cellcolor{green} 2.8 & \cellcolor{green} 4.3 & 6.6 & 7.5 & \cellcolor{green} 3.8 & \cellcolor{green} 5.9 & \cellcolor{green} 2.7 & \cellcolor{green} 4.9 & 8.0 \\
    & 2000 & \cellcolor{green} 5.9 & \cellcolor{green} 5.0 & \cellcolor{green} 4.2 & \cellcolor{green} 3.9 & \cellcolor{green} 5.0 & \cellcolor{green} 5.0 & \cellcolor{green} 4.9 & \cellcolor{green} 5.4 & \cellcolor{green} 4.1 & \cellcolor{green} 3.5 & \cellcolor{green} 5.3 & \cellcolor{green} 5.3 \\ \addlinespace[3pt]
    t5 & 250 & \cellcolor{red} 4.3 & \cellcolor{red} 4.1 & \cellcolor{red} 6.9 & \cellcolor{red} 3.1 & \cellcolor{red} 6.4 & \cellcolor{red} 6.4 & \cellcolor{red} 5.9 & \cellcolor{pink} 17.7 & \cellcolor{pink} 10.7 & \cellcolor{red} 8.3 & \cellcolor{pink} 17.7 & 32.4 \\
    & 500 & \cellcolor{red} 6.0 & \cellcolor{red} 5.2 & \cellcolor{red} 6.5 & \cellcolor{red} 4.5 & \cellcolor{red} 7.4 & \cellcolor{pink} 11.3 & \cellcolor{red} 9.5 & \cellcolor{pink} 22.4 & \cellcolor{pink} 22.8 & \cellcolor{pink} 13.4 & 33.9 & 33.9 \\
    & 1000 & \cellcolor{red} 4.9 & \cellcolor{red} 6.9 & \cellcolor{red} 5.2 & \cellcolor{red} 5.7 & \cellcolor{red} 8.0 & \cellcolor{pink} 10.8 & \cellcolor{pink} 17.7 & 33.0 & 33.1 & 33.0 & 42.7 & 52.7 \\
    & 2000 & \cellcolor{red} 6.0 & \cellcolor{red} 7.3 & \cellcolor{red} 5.8 & \cellcolor{red} 8.3 & \cellcolor{pink} 10.7 & \cellcolor{pink} 10.7 & 45.3 & 59.9 & 52.7 & 59.9 & 66.7 & 66.7 \\ \addlinespace[3pt]
    t3 & 250 & \cellcolor{red} 9.7 & \cellcolor{red} 3.6 & \cellcolor{pink} 10.3 & \cellcolor{red} 0.8 & \cellcolor{red} 2.0 & \cellcolor{red} 2.0 & \cellcolor{red} 5.6 & \cellcolor{pink} 13.5 & \cellcolor{red} 9.2 & \cellcolor{red} 6.0 & \cellcolor{pink} 13.5 & \cellcolor{pink} 26.9 \\
    & 500 & \cellcolor{pink} 15.8 & \cellcolor{red} 4.8 & \cellcolor{red} 9.5 & \cellcolor{red} 0.6 & \cellcolor{red} 1.3 & \cellcolor{red} 2.6 & \cellcolor{red} 7.8 & \cellcolor{pink} 16.2 & \cellcolor{pink} 16.9 & \cellcolor{red} 9.3 & \cellcolor{pink} 25.4 & \cellcolor{pink} 25.4 \\
    & 1000 & \cellcolor{pink} 14.2 & \cellcolor{red} 9.9 & \cellcolor{red} 9.7 & \cellcolor{red} 0.4 & \cellcolor{red} 0.6 & \cellcolor{red} 1.0 & \cellcolor{pink} 11.0 & \cellcolor{pink} 22.3 & \cellcolor{pink} 22.5 & \cellcolor{pink} 22.2 & 30.5 & 40.5 \\
    & 2000 & \cellcolor{pink} 25.9 & \cellcolor{pink} 16.6 & \cellcolor{pink} 16.5 & \cellcolor{red} 0.2 & \cellcolor{red} 0.3 & \cellcolor{red} 0.3 & \cellcolor{pink} 27.6 & 41.4 & 34.2 & 41.3 & 48.8 & 48.8 \\ \addlinespace[3pt]
    st3 & 250 & \cellcolor{red} 4.4 & \cellcolor{red} 5.4 & \cellcolor{red} 8.0 & \cellcolor{red} 4.5 & \cellcolor{red} 8.6 & \cellcolor{red} 8.6 & \cellcolor{pink} 10.4 & 31.2 & \cellcolor{pink} 19.2 & \cellcolor{pink} 18.3 & 31.2 & 49.0 \\
    & 500 & \cellcolor{red} 6.0 & \cellcolor{red} 6.9 & \cellcolor{red} 7.9 & \cellcolor{red} 6.3 & \cellcolor{pink} 10.1 & \cellcolor{pink} 14.7 & \cellcolor{pink} 22.4 & 44.2 & 44.3 & 31.9 & 57.2 & 57.2 \\
    & 1000 & \cellcolor{red} 5.5 & \cellcolor{red} 9.5 & \cellcolor{red} 6.9 & \cellcolor{red} 9.0 & \cellcolor{pink} 12.3 & \cellcolor{pink} 16.3 & 48.6 & 66.2 & 66.2 & 66.2 & \cellcolor{green} 74.7 & \cellcolor{green} 82.4 \\
    & 2000 & \cellcolor{red} 8.4 & \cellcolor{pink} 12.2 & \cellcolor{red} 9.8 & \cellcolor{pink} 14.6 & \cellcolor{pink} 17.9 & \cellcolor{pink} 17.9 & \cellcolor{green} 86.6 & \cellcolor{green} 92.9 & \cellcolor{green} 90.1 & \cellcolor{green} 92.9 & \cellcolor{green} 95.0 & \cellcolor{green} 95.0 \\
    \bottomrule
  \end{tabular}
  \caption{Estimated size and power of three different types of binomial test
(Wald, score, likelihood-ratio test (LRT)) applied to exceptions \\of the 97.5\%
and 99\% VaR estimates. Both one-sided and two-sided tests have been carried out.
Results are based on 10000 replications}
  \label{table:binomial}
\end{sidewaystable}

\subsubsection{Binomial test results}\label{sec:binom-test-results}

Table~\ref{table:binomial} shows the results for one-sided and two-sided binomial tests for the number of VaR exceptions at the 97.5\% and 99\% levels. In this table and in Table~\ref{table:multinomial} the following colour coding is used: green indicates good results ($\le 6\%$ for the size; $\ge 70\%$ for the power); red indicates poor results ($\ge 9\%$ for the size;  $\le 30\%$ for the power); dark red indicates very poor results ($\ge 12\%$ for the size; $\leq 10\%$ for the power).

\paragraph{97.5\% level.}

The size of the tests is generally reasonable. The score test in particular always seems to have a good size for all the different sample sizes in both the one-sided and two-sided tests.

The power of all the tests in extremely weak, which reflects the fact that the 97.5\% VaR values in all of the distributions are quite similar. Note that the one-sided tests are slightly more powerful at detecting the t5 and skewed t3 models whereas two-sided tests are slightly better at detecting the t3 model; the latter observation is due to the fact that the 97.5\% quantile of a (scaled) t3 is actually smaller than that of a normal distribution; see Table~\ref{table:rmvalues}.

\paragraph{99\% level.}

At this level the size is more variable and it is often too high in the smaller samples; in particular, the one-sided LRT (the Basel exception test) has a poor size in the case of the smallest sample. Once again the score test seems to have the best size properties.

The tests are more powerful in this case because there are more pronounced differences between the quantiles of the four models. One-sided tests are somewhat more powerful than two-sided tests since the non-normal models yield too many exceptions in comparison with the normal. The score test and LRT seem to be a little more powerful than the Wald test. Only in the case of the largest samples (1000 and 2000) from the distribution with the longest right tail (skewed t3) do we actually get high power (green cells).

\subsubsection{Multinomial test results}

\begin{figure}[h]
\caption{\label{fig:static-1-power} Size (first row) and power of the three multinomial tests as a function of $N$ The columns correspond to different sample sizes $n$ and the rows to the different underlying distributions $G$.}
\begin{center}
\includegraphics[width=15.5cm,height=15cm]{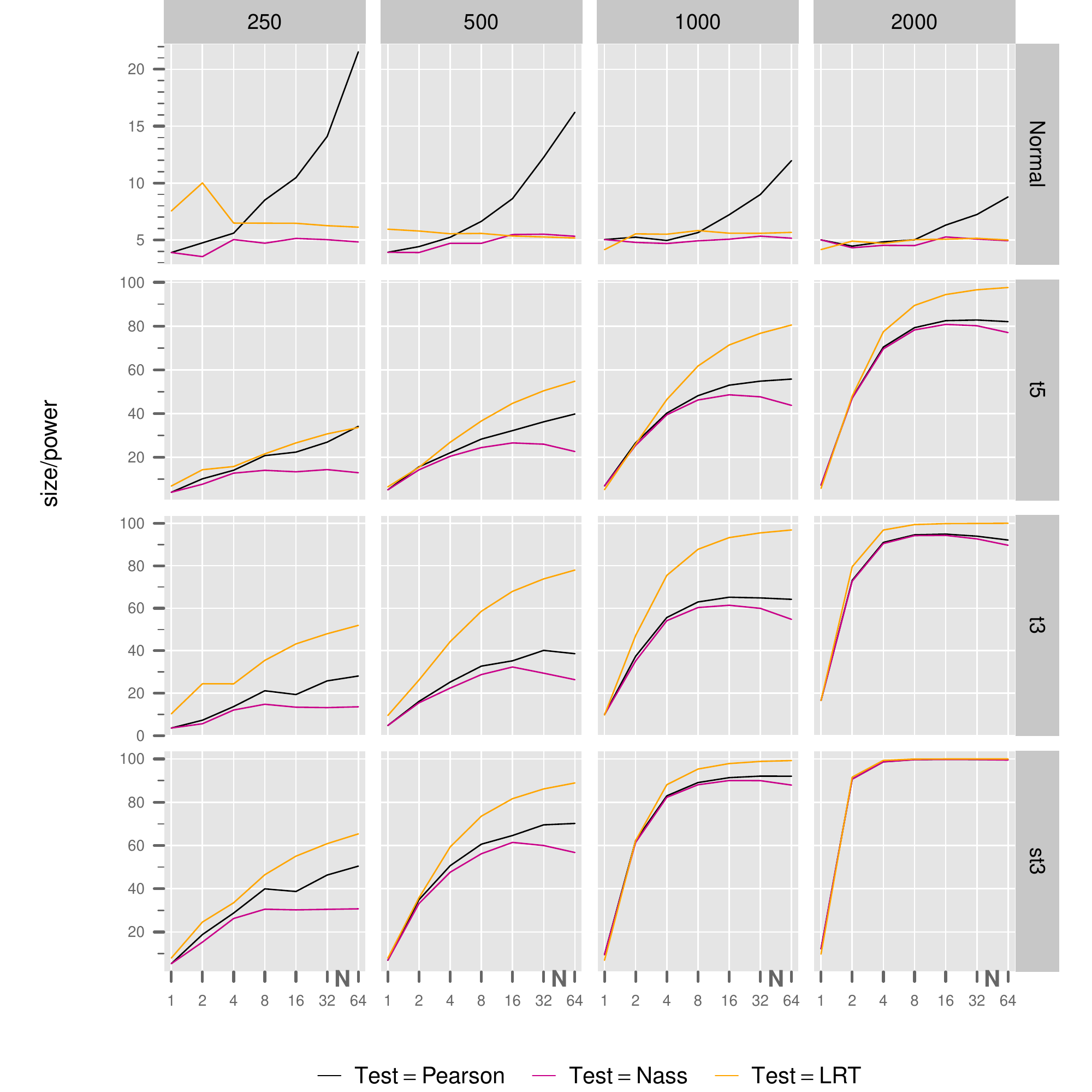}
\end{center}
\end{figure}

The results are shown in Table~\ref{table:multinomial} and displayed graphically in Figure~\ref{fig:static-1-power}. Note that, as discussed in Section~\ref{sec:case-n=1}, the Pearson test with $N=1$ gives identical results to the two-sided score test in Table~\ref{table:binomial}. In the case $N=1$ the Nass statistic is very close to the value of the Pearson statistic and also gives much the same results. The LRT with $N=1$ is the two-sided LRT from Table~\ref{table:binomial}.

\begin{sidewaystable}[htbp]
\setlength{\tabcolsep}{1pt}
  \centering
  \begin{tabular}{*{2}{l}*{21}{r}}
    \toprule
     & test & \multicolumn{7}{c|}{Pearson}  & \multicolumn{7}{c |}{Nass} & \multicolumn{7}{c}{LRT} \\
    \cmidrule(lr){3-9} \cmidrule(lr){10-16} \cmidrule(lr){17-23}
    \( G \) & \( n \) \textbar\ \( N \) & \multicolumn{1}{c}{1} & \multicolumn{1}{c}{2} & \multicolumn{1}{c}{4} & \multicolumn{1}{c}{8} & \multicolumn{1}{c}{16} & \multicolumn{1}{c}{32} & \multicolumn{1}{c|}{64} & \multicolumn{1}{c}{1} & \multicolumn{1}{c}{2} & \multicolumn{1}{c}{4} & \multicolumn{1}{c}{8} & \multicolumn{1}{c}{16} & \multicolumn{1}{c}{32} & \multicolumn{1}{c |}{64} & \multicolumn{1}{c}{1} & \multicolumn{1}{c}{2} & \multicolumn{1}{c}{4} & \multicolumn{1}{c}{8} & \multicolumn{1}{c}{16} & \multicolumn{1}{c}{32} & \multicolumn{1}{c}{64} \\
    \midrule
    Normal & 250 & \cellcolor{green} 3.9 & \cellcolor{green} 4.7 & \cellcolor{green} 5.6 & 8.5 & \cellcolor{pink} 10.5 & \cellcolor{red} 14.1 & \cellcolor{red} 21.5 & \cellcolor{green} 3.9 & \cellcolor{green} 3.5 & \cellcolor{green} 5.0 & \cellcolor{green} 4.7 & \cellcolor{green} 5.1 & \cellcolor{green} 5.0 & \cellcolor{green} 4.8 & 7.5 & \cellcolor{pink} 10.0 & 6.5 & 6.5 & 6.5 & 6.2 & 6.1 \\
    & 500 & \cellcolor{green} 3.9 & \cellcolor{green} 4.4 & \cellcolor{green} 5.2 & 6.6 & 8.6 & \cellcolor{red} 12.3 & \cellcolor{red} 16.2 & \cellcolor{green} 3.9 & \cellcolor{green} 3.9 & \cellcolor{green} 4.7 & \cellcolor{green} 4.7 & \cellcolor{green} 5.5 & \cellcolor{green} 5.5 & \cellcolor{green} 5.3 & \cellcolor{green} 5.9 & \cellcolor{green} 5.8 & \cellcolor{green} 5.5 & \cellcolor{green} 5.6 & \cellcolor{green} 5.3 & \cellcolor{green} 5.3 & \cellcolor{green} 5.2 \\
    & 1000 & \cellcolor{green} 5.0 & \cellcolor{green} 5.2 & \cellcolor{green} 5.0 & \cellcolor{green} 5.6 & 7.2 & 9.0 & \cellcolor{pink} 12.0 & \cellcolor{green} 5.0 & \cellcolor{green} 4.8 & \cellcolor{green} 4.7 & \cellcolor{green} 4.9 & \cellcolor{green} 5.1 & \cellcolor{green} 5.3 & \cellcolor{green} 5.1 & \cellcolor{green} 4.1 & \cellcolor{green} 5.5 & \cellcolor{green} 5.5 & \cellcolor{green} 5.8 & \cellcolor{green} 5.6 & \cellcolor{green} 5.6 & \cellcolor{green} 5.7 \\
    & 2000 & \cellcolor{green} 5.0 & \cellcolor{green} 4.5 & \cellcolor{green} 4.8 & \cellcolor{green} 5.0 & 6.3 & 7.2 & 8.8 & \cellcolor{green} 5.0 & \cellcolor{green} 4.3 & \cellcolor{green} 4.5 & \cellcolor{green} 4.5 & \cellcolor{green} 5.3 & \cellcolor{green} 5.1 & \cellcolor{green} 4.9 & \cellcolor{green} 4.2 & \cellcolor{green} 4.9 & \cellcolor{green} 4.7 & \cellcolor{green} 5.0 & \cellcolor{green} 5.1 & \cellcolor{green} 5.1 & \cellcolor{green} 5.0 \\ \addlinespace[3pt]
    t5 & 250 & \cellcolor{red} 4.1 & \cellcolor{pink} 10.2 & \cellcolor{pink} 14.1 & \cellcolor{pink} 20.8 & \cellcolor{pink} 22.4 & \cellcolor{pink} 27.0 & 34.2 & \cellcolor{red} 4.1 & \cellcolor{red} 7.7 & \cellcolor{pink} 12.8 & \cellcolor{pink} 14.1 & \cellcolor{pink} 13.4 & \cellcolor{pink} 14.4 & \cellcolor{pink} 13.0 & \cellcolor{red} 6.9 & \cellcolor{pink} 14.4 & \cellcolor{pink} 15.8 & \cellcolor{pink} 21.6 & \cellcolor{pink} 26.6 & 30.7 & 33.7 \\
    & 500 & \cellcolor{red} 5.2 & \cellcolor{pink} 15.7 & \cellcolor{pink} 22.1 & \cellcolor{pink} 28.4 & 32.2 & 36.2 & 39.8 & \cellcolor{red} 5.2 & \cellcolor{pink} 14.3 & \cellcolor{pink} 20.5 & \cellcolor{pink} 24.5 & \cellcolor{pink} 26.6 & \cellcolor{pink} 26.0 & \cellcolor{pink} 22.7 & \cellcolor{red} 6.5 & \cellcolor{pink} 15.5 & \cellcolor{pink} 26.9 & 36.6 & 44.7 & 50.4 & 54.8 \\
    & 1000 & \cellcolor{red} 6.9 & \cellcolor{pink} 26.7 & 40.2 & 48.2 & 53.0 & 54.8 & 55.8 & \cellcolor{red} 6.9 & \cellcolor{pink} 25.5 & 39.5 & 46.2 & 48.6 & 47.7 & 43.8 & \cellcolor{red} 5.2 & \cellcolor{pink} 26.1 & 46.4 & 61.8 & \cellcolor{green} 71.4 & \cellcolor{green} 76.7 & \cellcolor{green} 80.5 \\
    & 2000 & \cellcolor{red} 7.3 & 47.2 & \cellcolor{green} 70.4 & \cellcolor{green} 79.3 & \cellcolor{green} 82.5 & \cellcolor{green} 82.8 & \cellcolor{green} 82.0 & \cellcolor{red} 7.3 & 47.0 & 69.6 & \cellcolor{green} 78.2 & \cellcolor{green} 80.8 & \cellcolor{green} 80.2 & \cellcolor{green} 77.0 & \cellcolor{red} 5.8 & 48.0 & \cellcolor{green} 77.4 & \cellcolor{green} 89.5 & \cellcolor{green} 94.4 & \cellcolor{green} 96.6 & \cellcolor{green} 97.6 \\ \addlinespace[3pt]
    t3 & 250 & \cellcolor{red} 3.6 & \cellcolor{red} 7.3 & \cellcolor{pink} 13.7 & \cellcolor{pink} 21.1 & \cellcolor{pink} 19.4 & \cellcolor{pink} 25.8 & \cellcolor{pink} 28.1 & \cellcolor{red} 3.6 & \cellcolor{red} 5.6 & \cellcolor{pink} 12.1 & \cellcolor{pink} 14.8 & \cellcolor{pink} 13.4 & \cellcolor{pink} 13.2 & \cellcolor{pink} 13.6 & \cellcolor{pink} 10.3 & \cellcolor{pink} 24.4 & \cellcolor{pink} 24.4 & 35.4 & 43.2 & 48.0 & 51.9 \\
    & 500 & \cellcolor{red} 4.8 & \cellcolor{pink} 16.1 & \cellcolor{pink} 25.2 & 32.7 & 35.2 & 40.1 & 38.6 & \cellcolor{red} 4.8 & \cellcolor{pink} 15.5 & \cellcolor{pink} 22.4 & \cellcolor{pink} 28.7 & 32.3 & \cellcolor{pink} 29.4 & \cellcolor{pink} 26.4 & \cellcolor{red} 9.5 & \cellcolor{pink} 26.2 & 44.2 & 58.6 & 67.9 & \cellcolor{green} 73.8 & \cellcolor{green} 78.0 \\
    & 1000 & \cellcolor{red} 9.9 & 37.4 & 55.6 & 62.9 & 65.2 & 64.8 & 64.2 & \cellcolor{red} 9.9 & 35.2 & 54.1 & 60.3 & 61.4 & 59.9 & 54.7 & \cellcolor{red} 9.7 & 47.2 & \cellcolor{green} 75.4 & \cellcolor{green} 87.7 & \cellcolor{green} 93.2 & \cellcolor{green} 95.5 & \cellcolor{green} 96.8 \\
    & 2000 & \cellcolor{pink} 16.6 & \cellcolor{green} 73.1 & \cellcolor{green} 91.0 & \cellcolor{green} 94.5 & \cellcolor{green} 94.9 & \cellcolor{green} 93.9 & \cellcolor{green} 92.1 & \cellcolor{pink} 16.6 & \cellcolor{green} 72.7 & \cellcolor{green} 90.5 & \cellcolor{green} 94.2 & \cellcolor{green} 94.3 & \cellcolor{green} 92.6 & \cellcolor{green} 89.6 & \cellcolor{pink} 16.5 & \cellcolor{green} 79.5 & \cellcolor{green} 96.8 & \cellcolor{green} 99.4 & \cellcolor{green} 99.8 & \cellcolor{green} 99.9 & \cellcolor{green} 100.0 \\ \addlinespace[3pt]
    st3 & 250 & \cellcolor{red} 5.4 & \cellcolor{pink} 18.9 & \cellcolor{pink} 28.8 & 40.0 & 38.7 & 46.3 & 50.5 & \cellcolor{red} 5.4 & \cellcolor{pink} 15.3 & \cellcolor{pink} 26.3 & 30.5 & 30.2 & 30.5 & 30.7 & \cellcolor{red} 8.0 & \cellcolor{pink} 24.6 & 33.5 & 46.5 & 55.1 & 60.8 & 65.4 \\
    & 500 & \cellcolor{red} 6.9 & 34.9 & 50.7 & 60.6 & 64.6 & 69.5 & \cellcolor{green} 70.2 & \cellcolor{red} 6.9 & 33.2 & 47.6 & 56.2 & 61.4 & 60.0 & 56.8 & \cellcolor{red} 7.9 & 35.9 & 59.3 & \cellcolor{green} 73.6 & \cellcolor{green} 81.6 & \cellcolor{green} 86.2 & \cellcolor{green} 88.9 \\
    & 1000 & \cellcolor{red} 9.5 & 62.3 & \cellcolor{green} 83.0 & \cellcolor{green} 89.1 & \cellcolor{green} 91.3 & \cellcolor{green} 92.1 & \cellcolor{green} 92.0 & \cellcolor{red} 9.5 & 61.4 & \cellcolor{green} 82.3 & \cellcolor{green} 88.1 & \cellcolor{green} 90.0 & \cellcolor{green} 90.0 & \cellcolor{green} 87.9 & \cellcolor{red} 6.9 & 62.3 & \cellcolor{green} 88.1 & \cellcolor{green} 95.3 & \cellcolor{green} 97.9 & \cellcolor{green} 98.9 & \cellcolor{green} 99.2 \\
    & 2000 & \cellcolor{pink} 12.2 & \cellcolor{green} 90.7 & \cellcolor{green} 98.7 & \cellcolor{green} 99.7 & \cellcolor{green} 99.8 & \cellcolor{green} 99.8 & \cellcolor{green} 99.7 & \cellcolor{pink} 12.2 & \cellcolor{green} 90.7 & \cellcolor{green} 98.6 & \cellcolor{green} 99.7 & \cellcolor{green} 99.7 & \cellcolor{green} 99.7 & \cellcolor{green} 99.5 & \cellcolor{red} 9.8 & \cellcolor{green} 91.6 & \cellcolor{green} 99.3 & \cellcolor{green} 99.9 & \cellcolor{green} 100.0 & \cellcolor{green} 100.0 & \cellcolor{green} 100.0 \\
    \bottomrule
  \end{tabular}
  \caption{Estimated size and power of three different types of multinomial test
(Pearson, Nass, likelihood-ratio test (LRT)) based on exceptions of $N$ levels.
Results are based on 10000 replications}
  \label{table:multinomial}
\end{sidewaystable}

\paragraph{Size of the tests.}

The results for the size of the three tests are summarized in the first panel of Table~\ref{table:multinomial} where $G$ is Normal and in the first row of pictures in Figure~\ref{fig:static-1-power}.

The following points can be made.

\begin{itemize}
\item The size of the Pearson $\chi^2$-test deteriorates rapidly for $N\ge 8$ showing that this test is very sensitive to bin size.
\item The Nass test has the best size properties being very stable for all choices of $N$ and all sample sizes. In contrast to the other tests, the size is always less than or equal to 5\% for $2 \le N \le 8$; there is a slight tendency for the size to increase above 5\% when $N$ exceeds 8.
\item The LRT is over-sized in the smallest sample of size $n=250$ but otherwise has a reasonable size for all choices of $N$. In comparison with Nass, the size is often larger, tending to be a little more than 5\% except when $n=2000$ and $N\leq 8$.
\end{itemize}

\paragraph{Power of the tests.}

In rows 2--4 of Figure~\ref{fig:static-1-power} the power of the three tests is shown as a function of $N$ for different true underlying distributions $G$. It can be seen that for all $N$ the LRT is generally the most powerful test. The power of the Nass test is generally slightly lower than that of the Pearson test; it often tends to reach a maximum for $N=8$ or $N=16$ and then fall away - this would appear to be the price of the correction of the size of the Pearson test which the Nass test undertakes. However it is usually preferable to use a Nass test with $N=8$ than a Pearson test with $N=4$. Some further observations are as follows.

\begin{itemize}
\item[$\bullet$] {\it Student $t5$} (second row). This is the strongest challenge for the three tests because the tail is less heavy than for Student $t3$ and there is no skewness. Conclusions are as follows:
\begin{itemize}
\item[-] for the Nass and Pearson tests we require $n=2000$ and $N\geq 4$ to obtain a power over 70\% (coloured green in tables);
\item[-] for the LRT a power over 70\% can be obtained with $n\geq 1000$ and $N\geq 16$, or $n=2000$ and $N\geq4$.
\end{itemize}
\item[$\bullet$] {\it Student $t3$} (third row): 
\begin{itemize}
\item[-] as expected, the power is greater than that obtained for $t5$;
\item[-] to have a power in excess of 70\%, we need to take $n=2000$ for the Pearson and Nass tests; for the LRT, we can take $n=1000$ and $N\ge 4$, or $n=500$ and $N\ge 32$.
\end{itemize}
\item[$\bullet$] {\it skewed Student $t3$} (fourth row).
Here, we obtain power greater than 70\% for all three tests for $n=1000$, whenever $N \ge 4$. This is due to the fact that the skewness pushes the tail strongly to the right hand side.
\end{itemize}

In general the Nass test with $N=4$ or $8$ seems to be a good compromise between an acceptable size and power and to be slightly preferable to the Pearson text with $N=4$; an argument may also be made for preferring the Nass test with $N=4$ to the Pearson test with $N=4$ since it is reassuring to use a test whose size property is more stable than Pearson even if the power is very slightly reduced. In comparison with Nass, the LRT with $N=4$ or $N=8$ is a little oversized but very powerful; it comes into its own for larger data samples (see the case $n=2000$). 

If obtaining power to reject bad models is the overriding concern, then the LRT with $N > 8$ is extremely effective but starts to violate the principle that our test should not be much more burdensome to perform than a binomial test. It seems clear that, regardless of the test chosen, we should pick $N\geq 4$ since the resulting tests are much more powerful than a binomial test or a joint test of only two VaR levels.

In Table~\ref{table:binmultcomparison} we collect the results of the one-sided binomial score test of exceptions of the 99\% VaR (the most powerful binomial test) together with results for the Pearson and Nass tests with $N=4$ and the LRT with $N=4$ and $N=8$. The outperformance of the multinomial tests is most apparent in sample sizes of $n\geq 500$. In summary we find that:
\begin{itemize}
\item
For $n=250$ the power of all tests is less than 30\% for the case of t5 with the maximum value given by the LRT with $N=8$. The latter is also the most powerful test in the case of t3, being the only one with a power greater than 30\%. 
\item
For $n=500$ the Nass and Pearson tests with $N=4$ provide higher values than the binomial for t3 and st3 but very slightly lower values for t5. The LRT with $N=4$ is more powerful than the binomial, Pearson and Nass tests in all cases and the LRT with $N=8$ is even more powerful.
\item
The clearest advantages of the multinomial test over the best binomial test are for the largest sample sizes $n=1000$ and $n=2000$.  In this case all multinomial tests have higher power than the binomial test.
\end{itemize}
It should also be noted that the results from binomial tests are much more sensitive to the choice of $\alpha$. We have seen in Table~\ref{table:binomial} and Table~\ref{table:multinomial} that their performance for $\alpha =0.975$ is very poor. The multinomial tests using a range of thresholds are much less sensitive to the exact choice of these thresholds, which makes them a more reliable type of test.

\begin{table}[htbp]
  \centering
  \begin{tabular}{*{2}{l}*{5}{r}}
    \toprule
    \( G \) & \( n \) \textbar\ test & \multicolumn{1}{c}{Bin (0.99)} & \multicolumn{1}{c}{Pearson (4)} & \multicolumn{1}{c}{Nass (4)} & \multicolumn{1}{c}{LRT (4)} & \multicolumn{1}{c}{LRT (8)} \\
    \midrule
    Normal & 250 & \cellcolor{green} 4.0 & \cellcolor{green} 5.6 & \cellcolor{green} 5.0 & 6.5 & 6.5 \\
    & 500 & \cellcolor{green} 3.7 & \cellcolor{green} 5.2 & \cellcolor{green} 4.7 & \cellcolor{green} 5.5 & \cellcolor{green} 5.6 \\
    & 1000 & \cellcolor{green} 3.8 & \cellcolor{green} 5.0 & \cellcolor{green} 4.7 & \cellcolor{green} 5.5 & \cellcolor{green} 5.8 \\
    & 2000 & \cellcolor{green} 5.4 & \cellcolor{green} 4.8 & \cellcolor{green} 4.5 & \cellcolor{green} 4.7 & \cellcolor{green} 5.0 \\ \addlinespace[3pt]
    t5 & 250 & \cellcolor{pink} 17.7 & \cellcolor{pink} 14.1 & \cellcolor{pink} 12.8 & \cellcolor{pink} 15.8 & \cellcolor{pink} 21.6 \\
    & 500 & \cellcolor{pink} 22.4 & \cellcolor{pink} 22.1 & \cellcolor{pink} 20.5 & \cellcolor{pink} 26.9 & 36.6 \\
    & 1000 & 33.0 & 40.2 & 39.5 & 46.4 & 61.8 \\
    & 2000 & 59.9 & \cellcolor{green} 70.4 & 69.6 & \cellcolor{green} 77.4 & \cellcolor{green} 89.5 \\ \addlinespace[3pt]
    t3 & 250 & \cellcolor{pink} 13.5 & \cellcolor{pink} 13.7 & \cellcolor{pink} 12.1 & \cellcolor{pink} 24.4 & 35.4 \\
    & 500 & \cellcolor{pink} 16.2 & \cellcolor{pink} 25.2 & \cellcolor{pink} 22.4 & 44.2 & 58.6 \\
    & 1000 & \cellcolor{pink} 22.3 & 55.6 & 54.1 & \cellcolor{green} 75.4 & \cellcolor{green} 87.7 \\
    & 2000 & 41.4 & \cellcolor{green} 91.0 & \cellcolor{green} 90.5 & \cellcolor{green} 96.8 & \cellcolor{green} 99.4 \\ \addlinespace[3pt]
    st3 & 250 & 31.2 & \cellcolor{pink} 28.8 & \cellcolor{pink} 26.3 & 33.5 & 46.5 \\
    & 500 & 44.2 & 50.7 & 47.6 & 59.3 & \cellcolor{green} 73.6 \\
    & 1000 & 66.2 & \cellcolor{green} 83.0 & \cellcolor{green} 82.3 & \cellcolor{green} 88.1 & \cellcolor{green} 95.3 \\
    & 2000 & \cellcolor{green} 92.9 & \cellcolor{green} 98.7 & \cellcolor{green} 98.6 & \cellcolor{green} 99.3 & \cellcolor{green} 99.9 \\
    \bottomrule
  \end{tabular}
  \caption{Comparison of estimated size and power of one-sided binomial score test with $\alpha=0.99$ and Pearson, 
Nass and likelihood-ratio test with $N=4$ and LRT with $N=8$. 
Results are based on 10000 replications}
  \label{table:binmultcomparison}
\end{table}

\subsection{Static backtesting experiment}\label{sec:stat-backt-exper}

The style of backtest we implement (both here and in Section~\ref{sec:dyn}) is designed to mimic the procedure used in practice where models are continually updated to use the latest market data. We assume that the estimated model is updated every 10 steps; if these steps are interpreted as trading days this would correspond to every two trading week.
\subsubsection{Experimental design}
In each experiment we generate a total dataset of $n+n_2$ values from the true distribution $G$; we use the same four choices as in the previous section. The length $n$ of the backtest is fixed at the value 1000.

The modeller uses a rolling window of $n_2$ values to obtain an estimated distribution $F$, $n_2$ taking the values 250 and 500. We consider 4 possibilities for $F$: 
\begin{description}
\item[The oracle] who knows the correct distribution and its exact parameter values.
\item[The good modeller] who estimates the correct type of distribution (normal when $G$ is normal, Student t when $G$ is t5 or t3, skewed Student when $G$ is st3).
\item[The poor modeller] who always estimates a normal distribution (which is satisfactory only when $G$ is normal).
\item[The industry modeller] who uses the empirical distribution function by forming standard empirical quantile estimates, a method known as historical simulation in industry.
\end{description}

To make the rolling estimation procedure clear, the modellers begin by using the data $L_1,\ldots,L_{n_2}$ to form their model $F$ and make quantile estimates $\VaR_{\alpha_j,n_2+1}$ for $j=1,\ldots,N$. These are then compared with the true losses $\{L_{n_2+i},i=1,\ldots,10\}$ and the exceptions of each VaR level are counted. The modellers then roll the dataset forward 10 steps and use the data $L_{11},\ldots,L_{n_2+10}$ to make quantile estimates $\VaR_{\alpha_j,n_2+11}$ which are compared with the losses $\{L_{n_2+10+i},i=1,\ldots,10\}$; in total the models are thus re-estimated $n/10 =100$ times.

We consider the same three multinomial tests as before and the same numbers of levels $N$. The experiment is repeated $1000$ times to determine rejection rates.

\begin{sidewaystable}[htbp]
\setlength{\tabcolsep}{2pt}
\footnotesize
  \centering
  \begin{tabular}{*{3}{l}*{21}{r}}
    \toprule
     &  & Test & \multicolumn{7}{c|}{Pearson} & \multicolumn{7}{c|}{Nass} & \multicolumn{7}{c}{LRT} \\
    \cmidrule(lr){4-10} \cmidrule(lr){11-17} \cmidrule(lr){18-24}
    \( n_{2} \) & \( G \) & \( F \) \textbar\ \( N \) & \multicolumn{1}{c}{1} & \multicolumn{1}{c}{2} & \multicolumn{1}{c}{4} & \multicolumn{1}{c}{8} & \multicolumn{1}{c}{16} & \multicolumn{1}{c}{32} & \multicolumn{1}{c|}{64} & \multicolumn{1}{c}{1} & \multicolumn{1}{c}{2} & \multicolumn{1}{c}{4} & \multicolumn{1}{c}{8} & \multicolumn{1}{c}{16} & \multicolumn{1}{c}{32} & \multicolumn{1}{c|}{64} & \multicolumn{1}{c}{1} & \multicolumn{1}{c}{2} & \multicolumn{1}{c}{4} & \multicolumn{1}{c}{8} & \multicolumn{1}{c}{16} & \multicolumn{1}{c}{32} & \multicolumn{1}{c}{64} \\
    \midrule
    250 & Normal & Oracle & \cellcolor{green} 4.3 & \cellcolor{green} 5.6 & \cellcolor{green} 5.8 & \cellcolor{pink} 7.2 & \cellcolor{pink} 6.8 & \cellcolor{pink} 8.1 & \cellcolor{pink} 11.1 & \cellcolor{green} 4.3 & \cellcolor{green} 5.5 & \cellcolor{green} 5.2 & \cellcolor{pink} 6.8 & \cellcolor{green} 5.1 & \cellcolor{green} 5.0 & \cellcolor{green} 4.6 & \cellcolor{green} 3.8 & \cellcolor{pink} 6.3 & \cellcolor{green} 5.3 & \cellcolor{green} 5.3 & \cellcolor{green} 5.0 & \cellcolor{green} 5.3 & \cellcolor{green} 5.1 \\
    &  & Good & \cellcolor{green} 1.8 & \cellcolor{green} 3.5 & \cellcolor{green} 4.3 & \cellcolor{green} 5.8 & \cellcolor{pink} 7.5 & \cellcolor{pink} 11.0 & \cellcolor{red} 15.4 & \cellcolor{green} 1.8 & \cellcolor{green} 3.3 & \cellcolor{green} 4.1 & \cellcolor{green} 4.6 & \cellcolor{green} 5.5 & \cellcolor{pink} 7.0 & \cellcolor{green} 4.7 & \cellcolor{green} 1.4 & \cellcolor{green} 3.3 & \cellcolor{green} 2.8 & \cellcolor{green} 2.9 & \cellcolor{green} 2.8 & \cellcolor{green} 2.9 & \cellcolor{green} 3.0 \\
    &  & Poor & \cellcolor{black} NA & \cellcolor{black} NA & \cellcolor{black} NA & \cellcolor{black} NA & \cellcolor{black} NA & \cellcolor{black} NA & \cellcolor{black} NA & \cellcolor{black} NA & \cellcolor{black} NA & \cellcolor{black} NA & \cellcolor{black} NA & \cellcolor{black} NA & \cellcolor{black} NA & \cellcolor{black} NA & \cellcolor{black} NA & \cellcolor{black} NA & \cellcolor{black} NA & \cellcolor{black} NA & \cellcolor{black} NA & \cellcolor{black} NA & \cellcolor{black} NA \\
    &  & Industry & \cellcolor{red} 2.9 & \cellcolor{red} 6.2 & \cellcolor{red} 9.1 & \cellcolor{pink} 17.3 & 35.0 & 55.2 & \cellcolor{green} 72.7 & \cellcolor{red} 2.9 & \cellcolor{red} 5.5 & \cellcolor{red} 8.7 & \cellcolor{pink} 15.7 & \cellcolor{pink} 28.0 & 43.8 & 55.7 & \cellcolor{red} 1.9 & \cellcolor{red} 4.1 & \cellcolor{red} 6.9 & \cellcolor{pink} 15.2 & \cellcolor{pink} 29.4 & 49.4 & 68.0 \\ \addlinespace[3pt]
    & t5 & Oracle & \cellcolor{green} 5.6 & \cellcolor{green} 4.6 & \cellcolor{green} 5.0 & \cellcolor{pink} 6.2 & \cellcolor{pink} 7.3 & \cellcolor{pink} 9.4 & \cellcolor{red} 12.9 & \cellcolor{green} 5.6 & \cellcolor{green} 4.2 & \cellcolor{green} 4.9 & \cellcolor{green} 5.5 & \cellcolor{green} 4.9 & \cellcolor{green} 5.5 & \cellcolor{green} 4.7 & \cellcolor{green} 4.2 & \cellcolor{green} 5.4 & \cellcolor{pink} 6.1 & \cellcolor{green} 5.7 & \cellcolor{green} 5.8 & \cellcolor{green} 5.7 & \cellcolor{green} 5.5 \\
    &  & Good & \cellcolor{green} 2.3 & \cellcolor{green} 3.2 & \cellcolor{green} 4.3 & \cellcolor{green} 5.4 & \cellcolor{pink} 7.2 & \cellcolor{pink} 9.2 & \cellcolor{red} 13.8 & \cellcolor{green} 2.3 & \cellcolor{green} 2.9 & \cellcolor{green} 3.9 & \cellcolor{green} 4.2 & \cellcolor{green} 5.3 & \cellcolor{green} 5.6 & \cellcolor{green} 4.7 & \cellcolor{green} 1.2 & \cellcolor{green} 3.0 & \cellcolor{green} 3.4 & \cellcolor{green} 3.2 & \cellcolor{green} 3.8 & \cellcolor{green} 4.4 & \cellcolor{green} 5.1 \\
    &  & Poor & \cellcolor{red} 5.1 & 31.5 & 51.3 & 60.2 & 65.7 & 68.5 & \cellcolor{green} 70.3 & \cellcolor{red} 5.1 & 30.8 & 50.2 & 58.1 & 61.2 & 61.0 & 55.2 & \cellcolor{red} 3.1 & \cellcolor{pink} 28.7 & 54.4 & \cellcolor{green} 70.5 & \cellcolor{green} 80.0 & \cellcolor{green} 85.6 & \cellcolor{green} 89.1 \\
    &  & Industry & \cellcolor{red} 1.6 & \cellcolor{red} 5.0 & \cellcolor{red} 9.5 & \cellcolor{pink} 15.5 & 30.4 & 54.2 & \cellcolor{green} 71.4 & \cellcolor{red} 1.6 & \cellcolor{red} 4.6 & \cellcolor{red} 8.8 & \cellcolor{pink} 13.2 & \cellcolor{pink} 24.0 & 42.0 & 55.6 & \cellcolor{red} 1.1 & \cellcolor{red} 3.9 & \cellcolor{red} 7.2 & \cellcolor{pink} 13.0 & \cellcolor{pink} 25.8 & 43.8 & 64.7 \\ \addlinespace[3pt]
    & t3 & Oracle & \cellcolor{green} 5.5 & \cellcolor{green} 5.0 & \cellcolor{green} 5.4 & \cellcolor{green} 5.0 & \cellcolor{pink} 7.1 & \cellcolor{pink} 7.8 & \cellcolor{pink} 10.8 & \cellcolor{green} 5.5 & \cellcolor{green} 4.4 & \cellcolor{green} 4.7 & \cellcolor{green} 4.3 & \cellcolor{green} 4.9 & \cellcolor{green} 4.2 & \cellcolor{green} 4.5 & \cellcolor{green} 4.4 & \cellcolor{green} 4.2 & \cellcolor{green} 4.4 & \cellcolor{green} 4.3 & \cellcolor{green} 3.9 & \cellcolor{green} 3.7 & \cellcolor{green} 4.0 \\
    &  & Good & \cellcolor{green} 3.2 & \cellcolor{green} 4.3 & \cellcolor{green} 5.4 & \cellcolor{green} 5.8 & \cellcolor{pink} 7.1 & \cellcolor{pink} 10.0 & \cellcolor{red} 12.5 & \cellcolor{green} 3.2 & \cellcolor{green} 3.7 & \cellcolor{green} 5.1 & \cellcolor{green} 4.9 & \cellcolor{green} 4.8 & \cellcolor{green} 4.4 & \cellcolor{green} 5.6 & \cellcolor{green} 1.7 & \cellcolor{green} 3.4 & \cellcolor{green} 3.6 & \cellcolor{green} 2.9 & \cellcolor{green} 3.4 & \cellcolor{green} 3.5 & \cellcolor{green} 4.2 \\
    &  & Poor & \cellcolor{red} 4.3 & 46.0 & \cellcolor{green} 71.8 & \cellcolor{green} 82.8 & \cellcolor{green} 87.2 & \cellcolor{green} 88.1 & \cellcolor{green} 86.8 & \cellcolor{red} 4.3 & 45.0 & \cellcolor{green} 71.2 & \cellcolor{green} 81.4 & \cellcolor{green} 85.0 & \cellcolor{green} 84.0 & \cellcolor{green} 81.4 & \cellcolor{red} 3.3 & 47.7 & \cellcolor{green} 81.6 & \cellcolor{green} 94.1 & \cellcolor{green} 97.6 & \cellcolor{green} 99.1 & \cellcolor{green} 99.7 \\
    &  & Industry & \cellcolor{red} 2.3 & \cellcolor{red} 5.4 & \cellcolor{red} 7.9 & \cellcolor{pink} 16.3 & 32.6 & 54.8 & \cellcolor{green} 72.6 & \cellcolor{red} 2.3 & \cellcolor{red} 4.8 & \cellcolor{red} 7.6 & \cellcolor{pink} 14.5 & \cellcolor{pink} 26.1 & 44.3 & 56.9 & \cellcolor{red} 1.3 & \cellcolor{red} 3.5 & \cellcolor{red} 6.1 & \cellcolor{pink} 11.7 & \cellcolor{pink} 24.6 & 45.3 & 64.4 \\ \addlinespace[3pt]
    & st3 & Oracle & \cellcolor{green} 4.4 & \cellcolor{green} 5.4 & \cellcolor{green} 5.2 & \cellcolor{pink} 6.2 & \cellcolor{pink} 7.3 & \cellcolor{pink} 10.1 & \cellcolor{red} 13.7 & \cellcolor{green} 4.4 & \cellcolor{green} 5.3 & \cellcolor{green} 5.2 & \cellcolor{green} 5.5 & \cellcolor{green} 5.6 & \cellcolor{green} 5.9 & \cellcolor{pink} 6.5 & \cellcolor{green} 3.6 & \cellcolor{pink} 6.2 & \cellcolor{green} 5.6 & \cellcolor{green} 4.8 & \cellcolor{green} 4.7 & \cellcolor{green} 5.0 & \cellcolor{green} 4.9 \\
    &  & Good & \cellcolor{green} 2.1 & \cellcolor{green} 3.8 & \cellcolor{green} 4.2 & \cellcolor{pink} 6.9 & \cellcolor{pink} 7.7 & \cellcolor{pink} 11.0 & \cellcolor{red} 16.8 & \cellcolor{green} 2.1 & \cellcolor{green} 3.6 & \cellcolor{green} 4.2 & \cellcolor{pink} 6.3 & \cellcolor{green} 5.3 & \cellcolor{pink} 7.1 & \cellcolor{pink} 7.0 & \cellcolor{green} 1.5 & \cellcolor{green} 2.8 & \cellcolor{green} 3.4 & \cellcolor{green} 3.2 & \cellcolor{green} 3.2 & \cellcolor{green} 3.8 & \cellcolor{green} 4.1 \\
    &  & Poor & 34.9 & \cellcolor{green} 86.5 & \cellcolor{green} 96.5 & \cellcolor{green} 98.6 & \cellcolor{green} 99.4 & \cellcolor{green} 99.3 & \cellcolor{green} 99.3 & 34.9 & \cellcolor{green} 85.6 & \cellcolor{green} 96.2 & \cellcolor{green} 98.3 & \cellcolor{green} 99.0 & \cellcolor{green} 99.0 & \cellcolor{green} 98.5 & \cellcolor{pink} 26.8 & \cellcolor{green} 83.2 & \cellcolor{green} 97.1 & \cellcolor{green} 99.4 & \cellcolor{green} 99.7 & \cellcolor{green} 99.8 & \cellcolor{green} 99.9 \\
    &  & Industry & \cellcolor{red} 4.0 & \cellcolor{red} 6.9 & \cellcolor{red} 8.9 & \cellcolor{pink} 16.1 & 32.3 & 56.7 & \cellcolor{green} 73.8 & \cellcolor{red} 4.0 & \cellcolor{red} 6.0 & \cellcolor{red} 8.5 & \cellcolor{pink} 13.5 & \cellcolor{pink} 24.8 & 44.7 & 57.6 & \cellcolor{red} 1.8 & \cellcolor{red} 4.5 & \cellcolor{red} 6.8 & \cellcolor{pink} 11.7 & \cellcolor{pink} 26.4 & 46.5 & 63.7 \\ \addlinespace[6pt]
    500 & Normal & Oracle & \cellcolor{green} 5.0 & \cellcolor{green} 4.5 & \cellcolor{green} 4.8 & \cellcolor{green} 5.6 & \cellcolor{pink} \cellcolor{green} 6.0 & \cellcolor{pink} 7.8 & \cellcolor{red} 13.0 & \cellcolor{green} 5.0 & \cellcolor{green} 3.9 & \cellcolor{green} 4.4 & \cellcolor{green} 5.0 & \cellcolor{green} 3.9 & \cellcolor{green} 4.4 & \cellcolor{green} 3.5 & \cellcolor{green} 3.1 & \cellcolor{green} 4.8 & \cellcolor{green} 4.8 & \cellcolor{green} 4.5 & \cellcolor{green} 4.1 & \cellcolor{green} 4.1 & \cellcolor{green} 4.1 \\
    &  & Good & \cellcolor{green} 2.3 & \cellcolor{green} 3.3 & \cellcolor{green} 4.4 & \cellcolor{green} 5.8 & \cellcolor{pink} 7.3 & \cellcolor{pink} 8.3 & \cellcolor{pink} 11.8 & \cellcolor{green} 2.3 & \cellcolor{green} 2.7 & \cellcolor{green} 4.2 & \cellcolor{green} 4.4 & \cellcolor{green} 4.9 & \cellcolor{green} 5.1 & \cellcolor{green} 3.6 & \cellcolor{green} 1.5 & \cellcolor{green} 2.7 & \cellcolor{green} 2.9 & \cellcolor{green} 2.1 & \cellcolor{green} 2.6 & \cellcolor{green} 3.0 & \cellcolor{green} 3.1 \\
    &  & Poor & \cellcolor{black} NA & \cellcolor{black} NA & \cellcolor{black} NA & \cellcolor{black} NA & \cellcolor{black} NA & \cellcolor{black} NA & \cellcolor{black} NA & \cellcolor{black} NA & \cellcolor{black} NA & \cellcolor{black} NA & \cellcolor{black} NA & \cellcolor{black} NA & \cellcolor{black} NA & \cellcolor{black} NA & \cellcolor{black} NA & \cellcolor{black} NA & \cellcolor{black} NA & \cellcolor{black} NA & \cellcolor{black} NA & \cellcolor{black} NA & \cellcolor{black} NA \\
    &  & Industry & \cellcolor{green} 1.1 & \cellcolor{green} 1.8 & \cellcolor{green} 2.6 & \cellcolor{green} 4.8 & \cellcolor{pink} 9.0 & \cellcolor{red} 14.4 & \cellcolor{red} 25.9 & \cellcolor{green} 1.1 & \cellcolor{green} 1.6 & \cellcolor{green} 2.6 & \cellcolor{green} 3.8 & \cellcolor{green} 5.3 & 8.3 & \cellcolor{red} 12.3 & \cellcolor{green} 0.5 & \cellcolor{green} 1.5 & \cellcolor{green} 1.2 & \cellcolor{green} 3.1 & \cellcolor{green} 4.5 & 8.5 & \cellcolor{red} 15.6 \\ \addlinespace[3pt]
    & t5 & Oracle & \cellcolor{pink} 6.1 & \cellcolor{green} 5.0 & \cellcolor{green} 5.2 & \cellcolor{green} 5.9 & \cellcolor{pink} 6.6 & \cellcolor{pink} 9.1 & \cellcolor{red} 12.2 & \cellcolor{pink} 6.1 & \cellcolor{green} 4.2 & \cellcolor{green} 5.1 & \cellcolor{green} 4.5 & \cellcolor{green} 4.2 & \cellcolor{green} 5.3 & \cellcolor{green} 5.0 & \cellcolor{green} 3.8 & \cellcolor{green} 4.9 & \cellcolor{green} 5.9 & \cellcolor{green} 5.7 & \cellcolor{green} 4.4 & \cellcolor{green} 5.1 & \cellcolor{green} 5.1 \\
    &  & Good & \cellcolor{green} 2.6 & \cellcolor{green} 3.1 & \cellcolor{green} 4.5 & \cellcolor{green} 5.6 & \cellcolor{pink} 7.0 & \cellcolor{pink} 9.6 & \cellcolor{pink} 10.9 & \cellcolor{green} 2.6 & \cellcolor{green} 3.0 & \cellcolor{green} 4.3 & \cellcolor{green} 4.7 & \cellcolor{green} 4.6 & \cellcolor{green} 4.3 & \cellcolor{green} 3.7 & \cellcolor{green} 1.6 & \cellcolor{green} 2.9 & \cellcolor{green} 2.5 & \cellcolor{green} 3.4 & \cellcolor{green} 2.5 & \cellcolor{green} 2.9 & \cellcolor{green} 2.7 \\
    &  & Poor & \cellcolor{red} 5.6 & \cellcolor{pink} 27.6 & 43.6 & 54.0 & 59.6 & 59.5 & 62.3 & \cellcolor{red} 5.6 & \cellcolor{pink} 26.2 & 42.7 & 51.4 & 54.1 & 53.3 & 49.0 & \cellcolor{red} 3.4 & \cellcolor{pink} 24.6 & 48.7 & 68.5 & \cellcolor{green} 76.5 & \cellcolor{green} 82.8 & \cellcolor{green} 85.1 \\
    &  & Industry & \cellcolor{green} 1.3 & \cellcolor{green} 2.3 & \cellcolor{green} 2.9 & \cellcolor{green} 5.5 & 8.0 & \cellcolor{red} 15.0 & \cellcolor{red} 24.2 & \cellcolor{green} 1.3 & \cellcolor{green} 2.1 & \cellcolor{green} 2.8 & \cellcolor{green} 4.8 & \cellcolor{green} 5.0 & \cellcolor{pink} 9.2 & \cellcolor{red} 12.9 & \cellcolor{green} 0.9 & \cellcolor{green} 1.9 & \cellcolor{green} 1.5 & \cellcolor{green} 2.8 & \cellcolor{green} 4.5 & \cellcolor{pink} 9.0 & \cellcolor{red} 16.8 \\ \addlinespace[3pt]
    & t3 & Oracle & \cellcolor{green} 5.9 & \cellcolor{green} 5.2 & \cellcolor{green} 4.6 & \cellcolor{green} 5.6 & \cellcolor{pink} 6.4 & \cellcolor{pink} 7.6 & \cellcolor{red} 12.0 & \cellcolor{green} 5.9 & \cellcolor{green} 4.9 & \cellcolor{green} 4.4 & \cellcolor{green} 3.7 & \cellcolor{green} 5.0 & \cellcolor{green} 4.7 & \cellcolor{green} 5.6 & \cellcolor{green} 4.1 & \cellcolor{green} 5.5 & \cellcolor{green} 5.0 & \cellcolor{green} 5.0 & \cellcolor{green} 4.7 & \cellcolor{green} 4.6 & \cellcolor{green} 4.0 \\
    &  & Good & \cellcolor{green} 3.0 & \cellcolor{green} 3.9 & \cellcolor{green} 4.0 & \cellcolor{green} 5.0 & \cellcolor{pink} 6.7 & \cellcolor{pink} 8.8 & \cellcolor{red} 12.0 & \cellcolor{green} 3.0 & \cellcolor{green} 3.6 & \cellcolor{green} 3.9 & \cellcolor{green} 4.2 & \cellcolor{green} 4.6 & \cellcolor{green} 4.8 & \cellcolor{green} 4.8 & \cellcolor{green} 2.4 & \cellcolor{green} 3.9 & \cellcolor{green} 3.9 & \cellcolor{green} 3.3 & \cellcolor{green} 2.4 & \cellcolor{green} 2.8 & \cellcolor{green} 2.9 \\
    &  & Poor & \cellcolor{red} 6.9 & 46.4 & \cellcolor{green} 70.5 & \cellcolor{green} 78.4 & \cellcolor{green} 79.9 & \cellcolor{green} 80.6 & \cellcolor{green} 81.4 & \cellcolor{red} 6.9 & 45.5 & 69.6 & \cellcolor{green} 76.3 & \cellcolor{green} 77.4 & \cellcolor{green} 77.1 & \cellcolor{green} 73.3 & \cellcolor{red} 6.2 & 51.6 & \cellcolor{green} 82.9 & \cellcolor{green} 92.8 & \cellcolor{green} 95.9 & \cellcolor{green} 97.3 & \cellcolor{green} 98.1 \\
    &  & Industry & \cellcolor{green} 1.8 & \cellcolor{green} 1.5 & \cellcolor{green} 2.6 & \cellcolor{green} 5.4 & 7.6 & \cellcolor{red} 14.7 & \cellcolor{red} 25.7 & \cellcolor{green} 1.8 & \cellcolor{green} 1.4 & \cellcolor{green} 2.6 & \cellcolor{green} 4.7 & \cellcolor{green} 5.4 & \cellcolor{pink} 9.2 & \cellcolor{red} 13.8 & \cellcolor{green} 0.7 & \cellcolor{green} 1.1 & \cellcolor{green} 1.8 & \cellcolor{green} 2.1 & \cellcolor{green} 3.7 & 8.0 & \cellcolor{red} 14.6 \\ \addlinespace[3pt]
    & st3 & Oracle & \cellcolor{green} 5.2 & \cellcolor{green} 4.1 & \cellcolor{green} 5.1 & \cellcolor{pink} \cellcolor{green} 6.0 & \cellcolor{pink} 7.6 & \cellcolor{pink} 9.2 & \cellcolor{red} 13.0 & \cellcolor{green} 5.2 & \cellcolor{green} 3.4 & \cellcolor{green} 4.8 & \cellcolor{green} 5.2 & \cellcolor{green} 5.3 & \cellcolor{pink} 6.2 & \cellcolor{pink} 6.2 & \cellcolor{green} 3.7 & \cellcolor{green} 4.8 & \cellcolor{green} 4.6 & \cellcolor{green} 4.3 & \cellcolor{green} 4.2 & \cellcolor{green} 4.3 & \cellcolor{green} 4.5 \\
    &  & Good & \cellcolor{green} 3.1 & \cellcolor{green} 3.7 & \cellcolor{green} 3.9 & \cellcolor{pink} \cellcolor{green} 6.0 & \cellcolor{pink} 8.0 & \cellcolor{pink} 10.0 & \cellcolor{red} 12.8 & \cellcolor{green} 3.1 & \cellcolor{green} 3.4 & \cellcolor{green} 3.6 & \cellcolor{green} 5.0 & \cellcolor{pink} \cellcolor{green} 6.0 & \cellcolor{green} 5.7 & \cellcolor{green} 4.8 & \cellcolor{green} 2.1 & \cellcolor{green} 2.5 & \cellcolor{green} 2.8 & \cellcolor{green} 2.8 & \cellcolor{green} 2.6 & \cellcolor{green} 2.4 & \cellcolor{green} 2.2 \\
    &  & Poor & \cellcolor{pink} 27.2 & \cellcolor{green} 81.1 & \cellcolor{green} 93.7 & \cellcolor{green} 96.7 & \cellcolor{green} 97.8 & \cellcolor{green} 98.5 & \cellcolor{green} 98.4 & \cellcolor{pink} 27.2 & \cellcolor{green} 80.2 & \cellcolor{green} 93.3 & \cellcolor{green} 96.4 & \cellcolor{green} 97.0 & \cellcolor{green} 97.9 & \cellcolor{green} 97.6 & \cellcolor{pink} 21.4 & \cellcolor{green} 79.4 & \cellcolor{green} 96.0 & \cellcolor{green} 99.3 & \cellcolor{green} 99.9 & \cellcolor{green} 100.0 & \cellcolor{green} 100.0 \\
    &  & Industry & \cellcolor{green} 1.5 & \cellcolor{green} 1.5 & \cellcolor{green} 2.8 & \cellcolor{green} 5.7 & 8.4 & \cellcolor{red} 15.4 & \cellcolor{red} 23.3 & \cellcolor{green} 1.5 & \cellcolor{green} 1.4 & \cellcolor{green} 2.5 & \cellcolor{green} 4.7 & \cellcolor{green} 5.4 & \cellcolor{pink} 9.2 & \cellcolor{pink} 10.8 & \cellcolor{green} 0.8 & \cellcolor{green} 1.1 & \cellcolor{green} 1.7 & \cellcolor{green} 2.6 & \cellcolor{green} 3.8 & 7.1 & \cellcolor{red} 14.1 \\
    \bottomrule
  \end{tabular}
  \caption{Rejection rates for various VaR estimation methods and various tests in the static backtesting experiment. Models are refitted after 10 simulated values and backtest length is 1000. Results are based on 1000 replications.}
  \label{table:backtest-static}
\end{sidewaystable}

\subsubsection{Results}

In Table~\ref{table:backtest-static} and again in Table~\ref{table:backtest-dynamic} we use the same colouring scheme as previously but a word of explanation is now required concerning the concepts of size and power.

The backtesting results for the oracle, who knows the correct model should clearly be judged in terms of size since we need to control the type one error of falsely rejecting the null hypothesis that the oracle's quantile ``estimates'' are accurate. We judge the results for the good modeller according to the same standards as the oracle . In doing this we make the judgement that a sample of size $n_2=250$ or $n_2 =500$ is sufficient to estimate quantiles parametrically in a static situation when a modeller chooses the right class of distribution. We would not want to have a high rejection rate that penalizes the good modeller too often in this situation. Thus we apply the size colouring scheme to both the oracle and the good modeller.

The backtesting results for the poor modeller should clearly be judged in terms of power. We want to obtain a high rejection rate for this modeller who is using the wrong distribution, regardless of how much data he or she is using. Hence the power colouring is applied in this case.

For the industry modeller the situation is more subtle. Empirical quantile estimation is an acceptable method provided that enough data is used. However it is less easy to say what is enough data because this depends on how heavy the tails of the underlying distribution are and how far into the tail the quantiles are estimated (which depends on $N$). To keep things simple we have made the arbitrary decision that a sample size of $n_2 =250$ is too small to permit the use of empirical quantile estimation and we have applied power colouring in this case; a modeller should be discouraged from using empirical quantile estimation in small samples.

On the other hand we have taken the view that $n_2=500$ is an acceptable sample size for empirical quantile estimation (particularly for $N$ values up to 4). We have applied size colouring in this case.

In general we are looking for a testing method that gives as much green colouring as possible in Table~\ref{table:backtest-static} and which minimizes the amount of red colouring.

The results for the oracle and the good modeller are in the desired green zone for all tests and all values of $N$ with the exception of the Pearson test with $N > 4$. It is in judging the results of the poor modeller that the increased power of the multinomial tests over the binomial test becomes apparent.
Indeed using a binomial test ($N=1$)
does not lead to an acceptable rejection rate for the poor modeller and using a test with $N=2$ is also generally insufficient, except for the skewed Student case. We infer that choosing a value $N\ge 4$ is necessary if we want to satisfy both criteria: a probability less than 6\% of rejecting the results of the modeller who uses the right model, as well as a probability below 30\%  (that is a power above 70\%) of accepting the results of the modeller who uses the wrong model.

Considering the different tests in more detail, the table shows that for the Pearson test, the best option is to set $N=4$; one might consider setting $N=8$ if $500$ values are used. Taking any more thresholds tends to lead to over-sized tests that reject the oracle more than should be the case and also reject the good modeller more often than a regulator might wish to. 

The Nass test is again very stable with respect to the choice of $N$: the size is mostly correct and the rejection rate for the good modeller is seldom above 6\% (except in certain cases for $n_2=250$ and $N\ge 32$).
 To obtain high power to reject the poor modeller, a choice of $N=4$ or $N=8$ seems reasonable and this leads to rejection rates that are comparable or superior to Pearson with similar values of $N$.

The LRT is also very stable with respect to size and to the rejection rate for the good modeller; we note that the sample size in Table~\ref{table:backtest-static} is always $n=1000$ and we only detected real issues with the size of the LRT in the smallest sample $n_2=250$ in Table~\ref{table:multinomial}. 
Moreover, for most values of $N$ we can obtain even higher power than Nass or Pearson for rejecting the bad results of the poor modeller. Note that for $n_2=250$, we need $N=8$ to reject the choice of a normal distribution with probability above 70\% when the true underlying distribution is Student $t5$, and we need $N=16$ when $n_2=1000$; the other tests are not able to attain this power.

In the case of the industry modeller, for a sample size $n=250$, the tests begin to expose the unreliability of the industry modeller for $N > 4$. This is to be expected because there are not enough points in the tail to permit accurate estimation of the more extreme quantiles. Ideally we want the industry modeller to be exposed in this situation so this is an argument for picking $N$ relatively high.

Increasing $n_2$ to 500 improves the situation for empirical quantile estimation, provided we do not consider too large a value of $N$. We obtain good green test results when setting the number of levels to be $N\le 8$ for the Pearson test and $N\le 16$ for the other  tests. Increasing $n_2$ further to, say, $n_2=1000$ (or four years of data) leads to a further reduction in the rejection rate for the industry modeller as empirical quantile estimation becomes an even more viable method of estimating the quantiles.

In summary, it is again clear that taking values of $N\geq 4$ gives reliable results which are superior to those obtained when $N=1$ or $N=2$. The use of only one or two quantile estimates does not seem sufficient to discriminate between light and heavy tails 
and a fortiori to construct an implicit backtest of expected shortfall based on $N$ VaR levels, a conclusion that has been noted in~\citet{bib:kratz-lok-mcneil-16}.

\subsection{Dynamic backtesting experiment} \label{sec:dyn}
Here the backtesting set-up is similar to that used in Section~\ref{sec:stat-backt-exper} but the experiment is conducted in a time-series setup. The true data-generating mechanism for the losses is a stationary GARCH model with Student innovations.

We choose to simulate data from a GARCH(1,1) model with Student t innovations; the parameters have been chosen by fitting this model to S\&P index log-returns for the period 2000--2012 (3389 values). The parameters of the GARCH equation in the standard notation are $\alpha_0 = 2.18\times 10^{-6}$, $\alpha_1 = 0.109$ and $\beta_1= 0.890$ while the degree of freedom of the Student innovation distribution is $\nu= 5.06$.

\subsubsection{Experimental design}

A variety of forecasters use different methods to estimate the conditional distribution of the losses at each time point and deliver VaR estimates. The length of the backtest is $n = 1000$ (approximately 4 years) as in Section~\ref{sec:stat-backt-exper} and each forecaster uses a rolling window of $n_2$ values to make their forecasts. We consider the values $n_2=500$ and $n_2=1000$; these window lengths are longer than in the static backtest study since more data is generally needed to estimate a GARCH model reliably. All models are re-estimated every 10 time steps. The experiment is repeated $500$ times to determine rejection rates for each forecaster. 

The different forecasting methods considered are listed below; for more details of the methodology, see~\citet{bib:mcneil-frey-embrechts-15}, Chapter 9.

\begin{description}
\item[Oracle:] the forecaster knows the correct model and its exact parameter values.
\item[GARCH.t:] the forecaster estimates the correct type of model (GARCH(1,1) with $t$ innovations). Note that he does not know the degree of freedom and has to estimate this parameter as well.
\item[GARCH.HS:] the forecaster uses a GARCH(1,1) model to estimate the dynamics of the losses but applies empirical quantile estimation to the residuals to estimate quantiles of the innovation distribution and hence quantiles of the conditional loss distribution; this method is often called filtered historical simulation in practice. We have already noted in the static backtesting experiment that empirical methods are only acceptable when we use a sufficient quantity of data.
\item[GARCH.EVT:] the forecaster uses a variant on GARCH.HS in which an EVT tail model is used to get slightly more accurate estimates of conditional quantiles in small samples.
\item[GARCH.norm:] the forecaster estimates a GARCH(1,1) model with normal innovation distribution.
\item[ARCH.t:] the forecaster misspecifies the dynamics of the losses by choosing an ARCH(1) model but correctly guesses that the innovations are $t$-distributed.
\item[ARCH.norm:] as in GARCH.norm but the forecaster misspecifies the dynamics to be ARCH(1). 
\item[HS:] the forecaster applies standard empirical quantile estimation to the data, the method used by the \emph{industry} modeller in Section~\ref{sec:stat-backt-exper}. As well as completely neglecting the dynamics of market losses, this method is prone to the drawbacks of empirical quantile estimation in small samples.
\end{description}

\subsubsection{Results}

We summarize the results found in Table~\ref{table:backtest-dynamic} considering first the true model (oracle), then the good models (GARCH.t, GARCH.HS, GARCH.EVT), and finally the poor models (GARCH.norm, the ARCH models and HS). Note that we will include GARCH.HS among the good models based on our assumption in the static experiment of Section~\ref{sec:stat-backt-exper}, that a data sample of size $n_2 =500$ is sufficient for empirical quantile estimation; this is clearly an arbitrary judgement.

We observe, in general, that the three tests are better able to discriminate between the tails of the models (heavy-tailed versus light-tailed) than between different forms of dynamics (GARCH versus ARCH). The binomial test ($N=1$) is unable to discriminate between the Student t and normal innovation distributions in the GARCH model; taking $N=2$ slightly improves the result, but we need $N\ge 4$ in order to really expose the deficiencies of the normal innovation distribution when the true process has heavier-tailed innovations. All tests are very powerful, for any choice of $N$, when both the choice of dynamics and the choice of innovation distribution are wrong (ARCH.norm).

\begin{sidewaystable}[htbp]
\setlength{\tabcolsep}{1pt}
\footnotesize
  \centering
  \begin{tabular}{*{3}{l}*{7}{c}*{7}{c}*{9}{c}}
    \toprule
  & Test & \multicolumn{7}{c |}{Pearson} & &\multicolumn{7}{c|}{Nass} & & \multicolumn{7}{c}{LRT} \\
    \cmidrule(lr){3-9} \cmidrule(lr){11-17} \cmidrule(lr){19-25} 
    \( n_{2} \) &  \( F \) \textbar\ \( N \) & \multicolumn{1}{c}{1} & \multicolumn{1}{c}{2} & \multicolumn{1}{c}{4} & \multicolumn{1}{c}{8} & \multicolumn{1}{c}{16} & \multicolumn{1}{c}{32} & \multicolumn{1}{c|}{64} & & \multicolumn{1}{c}{1} & \multicolumn{1}{c}{2} & \multicolumn{1}{c}{4} & \multicolumn{1}{c}{8} & \multicolumn{1}{c}{16} & \multicolumn{1}{c}{32} & \multicolumn{1}{c|}{64} && \multicolumn{1}{c}{1} & \multicolumn{1}{c}{2} & \multicolumn{1}{c}{4} & \multicolumn{1}{c}{8} & \multicolumn{1}{c}{16} & \multicolumn{1}{c}{32} & \multicolumn{1}{c}{64} \\
    \midrule
    500 & Oracle & \cellcolor{green}  6.0 & \cellcolor{green} 4.0 & \cellcolor{green} 3.8 & \cellcolor{green}  5.0 & \cellcolor{green}  5.6 & \cellcolor{pink}  9.6 &  \cellcolor{pink} 9.4 & & \cellcolor{green} 6.0 & \cellcolor{green} 3.2 & \cellcolor{green} 3.6 & \cellcolor{green} 4.2 & \cellcolor{green} 2.8 & \cellcolor{green} 4.8 & \cellcolor{green} 4.0 && \cellcolor{green} 3.4 & \cellcolor{green} 4.8 & \cellcolor{green}  5.2 & \cellcolor{green} 5.0 &  \cellcolor{green} 5.4 &  \cellcolor{green} 5.6 & \cellcolor{green} 5.6 \\
    & GARCH.t & 6.8 & \cellcolor{green} 5.6 & 6.2 & 8.0 & 8.2 & \cellcolor{red}12.8 &\cellcolor{red} 18.4 & & 6.8 & \cellcolor{green} 5.0 & \cellcolor{green} 6.0 & 6.2 & 7.0 & 7.6 & 6.4 & & \cellcolor{green} 4.6 & \cellcolor{green} 5.0 & \cellcolor{green} 5.4 & \cellcolor{green} 4.8 & \cellcolor{green} 4.6 & \cellcolor{green} 5.2 & \cellcolor{green} 5.2 \\
    & GARCH.HS & \cellcolor{green} 1.6 & \cellcolor{green} 1.6 & \cellcolor{green} 4.4 &  \cellcolor{pink} 11.8 &  \cellcolor{red} 25.4 &  \cellcolor{red} 92.0 &  \cellcolor{red} 98.8 &&
\cellcolor{green} 1.6 & \cellcolor{green} 1.4 & \cellcolor{green} 4.4 &  \cellcolor{pink} 10.8 &  \cellcolor{red} 20.4 &  \cellcolor{red} 85.0 &  \cellcolor{red} 97.4 && 
\cellcolor{green} 0.8 & \cellcolor{green} 1.6 & \cellcolor{green} 3.6 & \cellcolor{green} 2.0 & \cellcolor{green} 2.0 & \cellcolor{green} 5.6 &  \cellcolor{red} 13.0 \\
    & GARCH.EVT & \cellcolor{green} 2.2 & \cellcolor{green} 3.6 & \cellcolor{green} 3.6 & 7.2 & 7.6 &  \cellcolor{red} 12.2 &  \cellcolor{red} 16.2 & & \cellcolor{green} 2.2 & \cellcolor{green} 3.6 & \cellcolor{green} 3.2 & \cellcolor{green} 6.0 & \cellcolor{green} 5.0 & 6.8 & 7.4 & & \cellcolor{green} 0.8 & \cellcolor{green} 3.6 & \cellcolor{green} 2.0 & \cellcolor{green} 0.8 & \cellcolor{green} 1.2 & \cellcolor{green} 1.0 & \cellcolor{green} 1.0 \\
    & GARCH.norm &  \cellcolor{pink} 10.8 &  34.0 &   50.4 &  61.6 &  66.0 &  68.6 &  69.4 & & 
 \cellcolor{pink} 10.8 &  32.2 &   49.4 &  60.0 &  61.4 &   63.2 &  55.4 & &
 \cellcolor{red} 8.2 &   34.0 &   55.2 &  \cellcolor{green} 71.2 &  \cellcolor{green} 79.8 &  \cellcolor{green} 85.0 &  \cellcolor{green} 87.4 \\
    & ARCH.t &  34.0 &  32.4 &   32.0 &  \cellcolor{pink}  29.8 &  \cellcolor{pink} 29.4 &  33.8 &   39.6 & & 34.0 &  31.4 & 31.2 &  \cellcolor{pink}  28.6 &  \cellcolor{pink}  26.8 &  \cellcolor{pink}  25.6 &  \cellcolor{pink}  28.0 &  & 
 30.4 &   31.2 & 31.4 &  31.6 &  31.8 &  31.8 &  31.2 \\
    & ARCH.norm &  \cellcolor{green} 96.2 &  \cellcolor{green} 99.6 &  \cellcolor{green}99.6 &  \cellcolor{green} 99.8 &  \cellcolor{green} 100.0 &  \cellcolor{green} 100.0 &  \cellcolor{green} 100.0 & & \cellcolor{green} 96.2 &  \cellcolor{green} 99.6 & \cellcolor{green} 99.6 &  \cellcolor{green} 99.8 &  \cellcolor{green} 100.0 & \cellcolor{green} 100.0 & \cellcolor{green} 100.0 &&
 \cellcolor{green} 95.0 &  \cellcolor{green} 99.6 &  \cellcolor{green} 99.6 &  \cellcolor{green} 99.8 &  \cellcolor{green} 100.0 &  \cellcolor{green} 100.0 &  \cellcolor{green} 100.0 \\
    & HS &   39.4 &   38.8 &   39.8 &  42.2 &   49.8 &  \cellcolor{green} 80.2 &  \cellcolor{green} 90.0 &&   39.4 &  38.6 &   39.8 &  40.8 &  48.0 &  \cellcolor{green} 77.0 &  \cellcolor{green} 85.0 && 
  36.8 &    40.0 &   44.8 &   43.8 &  42.2 &  49.2 &   55.8 \\ \addlinespace[3pt]
    1000 & Oracle & \cellcolor{green} 4.2 & \cellcolor{green} 3.4 & \cellcolor{green} 3.8 & \cellcolor{green} 3.4 & \cellcolor{green} 5.0 & 7.6 & \cellcolor{pink}10.2 & &\cellcolor{green} 4.2 & \cellcolor{green} 3.2 & \cellcolor{green} 3.8 & \cellcolor{green} 2.8 & \cellcolor{green} 3.6 & \cellcolor{green} 3.8 & \cellcolor{green} 3.8 && \cellcolor{green} 3.4 & \cellcolor{green} 2.6 & \cellcolor{green} 3.2 & \cellcolor{green} 2.6 & \cellcolor{green} 2.6 & \cellcolor{green} 2.4 & \cellcolor{green} 2.4 \\
    & GARCH.t & \cellcolor{green} 5.8 & \cellcolor{green} 4.6 & 6.2 & \cellcolor{green} 5.2 & \cellcolor{green} 6.0 & \cellcolor{pink}11.2 & \cellcolor{red}12.8 && 
\cellcolor{green} 5.8 & \cellcolor{green} 3.8 & \cellcolor{green} 5.2 & \cellcolor{green} 3.2 & \cellcolor{green} 4.2 & 6.6 & 6.6 & &
\cellcolor{green} 4.4 & \cellcolor{green} 2.8 & \cellcolor{green} 3.6 & \cellcolor{green} 3.6 & \cellcolor{green} 3.4 & \cellcolor{green} 4.8 & \cellcolor{green} 4.0 \\
    & GARCH.HS & \cellcolor{green} 3.0 & \cellcolor{green} 2.0 & \cellcolor{green} 2.6 & \cellcolor{green} 4.4 &  \cellcolor{pink} 10.2 &  \cellcolor{red} 21.2 &  \cellcolor{red}  69.0 & &
\cellcolor{green} 3.0 & \cellcolor{green} 1.8 & \cellcolor{green} 2.2 & \cellcolor{green} 4.0 & 7.2 &  \cellcolor{red} 13.2 &  \cellcolor{red}  56.0 && 
\cellcolor{green} 1.8 & \cellcolor{green} 1.6 & \cellcolor{green} 2.6 & \cellcolor{green} 3.4 & \cellcolor{green} 3.8 & \cellcolor{green} 3.0 & \cellcolor{green} 5.2 \\
    & GARCH.EVT & \cellcolor{green} 2.6 & \cellcolor{green} 3.4 & \cellcolor{green} 4.2 & \cellcolor{green} 4.2 & 7.0 & 7.2 &  \cellcolor{pink}10.4 & &
\cellcolor{green} 2.6 & \cellcolor{green} 3.4 & \cellcolor{green} 3.6 & \cellcolor{green} 3.4 & \cellcolor{green} 5.0 & \cellcolor{green} 3.8 & \cellcolor{green} 4.2 && 
\cellcolor{green} 1.6 & \cellcolor{green} 4.6 & \cellcolor{green} 3.2 & \cellcolor{green} 2.6 & \cellcolor{green} 2.0 & \cellcolor{green} 1.8 & \cellcolor{green} 1.8 \\
    & GARCH.norm &  \cellcolor{red} 9.4 &  30.6 &    45.6 &  52.2 &  58.4 &   61.4 &   63.6 &&  \cellcolor{red} 9.4 &  \cellcolor{pink} 29.8 &  44.6 &  49.6 &   53.0 &   54.6 &  50.6 &&  
\cellcolor{red} 6.4 &  \cellcolor{pink} 28.4 &   49.8 &  65.2 &  \cellcolor{green} 76.6 &  \cellcolor{green} 83.0 &  \cellcolor{green} 86.6 \\
   & ARCH.t &   42.4 &   36.8 &   32.8 & \cellcolor{pink} 28.0 & \cellcolor{pink} 25.0 &  30.2 &  33.8 & & 
42.4 &   36.0 &  32.2 &  \cellcolor{pink} 27.0 &  \cellcolor{pink}23.0 &  \cellcolor{pink} 25.6 &  \cellcolor{pink} 27.2 &&  
 39.4 &  40.2 & 39.6 &  40.0 &  40.2 &   40.4 &  40.8 \\
    & ARCH.norm &  \cellcolor{green} 82.8 &  \cellcolor{green} 94.6 &   \cellcolor{green} 97.6 &   \cellcolor{green} 98.2 &   \cellcolor{green} 98.6 &  \cellcolor{green} 98.2 &  \cellcolor{green} 98.6 &&   \cellcolor{green} 82.8 &  \cellcolor{green} 94.4 &   \cellcolor{green} 97.6 &   \cellcolor{green} 98.0 &   \cellcolor{green} 98.2 &   \cellcolor{green} 98.0 &   \cellcolor{green} 97.8 & & 
 \cellcolor{green} 80.8 &   \cellcolor{green} 95.2 &   \cellcolor{green} 98.8 &   \cellcolor{green} 98.8 &   \cellcolor{green} 99.2 &   \cellcolor{green} 99.2 &   \cellcolor{green} 99.4 \\
    & HS &  51.4 &   51.0 &    45.0 &  37.2 &   34.6 &  39.8 & 55.6 & & 
51.4  &   50.6 &    44.4 &  35.0 &   31.8 &    36.2 & 49.8 & &  
  49.2 &  51.8 &    52.6 &   53.8 &  53.8 &  53.0 &   55.4 \\
    \bottomrule
  \end{tabular}
  \caption{Estimated size and power of three different types of multinomial test
(Pearson, Nass, likelihood-ratio test (LRT)) based on exceptions \\of $N$ levels.
Results are based on 500 replications of backtests of length 1000}
  \label{table:backtest-dynamic}
\end{sidewaystable}

\paragraph{Pearson test.}
In view of our previous experiments and the deterioration in the size properties of the Pearson test when $N$ is too large, it seems advisable for this test, to restrict attention to the case $N=4$ when $n_2=500$, or $N=8$ when $n_2=1000$. We see that the probability of rejecting the good modellers starts to increase for $N>4$ when $n_2=500$ and  for $N>8$ when $n_2=1000$.

The Pearson test discriminates better between innovation models than between the different dynamics. The power to reject a GARCH.norm (mispecified innovation distribution) increases with $N$ and gives reasonable results (no colouring) from $N=4$; note again that the binomial test ($N=1$) has very low power. For ARCH.t (mispecified dynamics), the power of the test is less than for GARCH.norm, but acceptable if $N<8$. For HS the power results are reasonable for any $N$, suggesting the test has some ability to detect modellers who neglect the modelling of dynamics. Overall, $N=4$ seems the best choice, with  reasonable power to reject poor forecasters using GARCH.norm, ARCH.t and HS.

Note that the results for the Pearson test with $N=4$ are broadly comparable for estimation windows of $n_2=500$ and $n_2 =1000$.

\paragraph{Nass test.} 

The colour coding shows that the Nass test exhibits a fairly similar pattern to the Pearson test, but with no deterioration in the size of the test for large $N$ ($N=32$ or $64$).

The Nass test has an acceptably low rejection rate for the good modeller for most values of $N$, and in the case of GARCH.HS, for $N<16$. In view of our previous discussion we focus on the cases $N=4$ and $N=8$.

For $N=4$, we obtain broadly comparable results to the Pearson test with $N=4$ with only a very slight reduction in the power to reject GARCH.norm, ARCH.t and HS. 

\paragraph{LRT.} 

This test tends to give very stable inference for all values of $N$; this can be seen by the fact that many rows of figures are uniformly green or uniformly white (with no deterioration for large $N$).
This stability helps to validate the results obtained for $N=4,8$, with the Pearson and Nass tests.

Considering first the good modellers and comparing with the Pearson and Nass tests, the LRT is slightly oversized when $n_2=500$ (as in the static case)  but has the best size when $n_2=1000$. It gives the best results for GARCH.t with a slight improvement from $n_2=500$ to 1000.
It gives stable and low rejection rates for GARCH.HS until $N=32$ for $n_2=500$ and until $N=64$ for $n_2 = 1000$; it also gives stable and low rejection rates for GARCH.EVT for any $N$.

Considering the poor modellers, the LRT has increasing power with $N$ for GARCH.norm; in contrast to the other tests the power exceeds the threshold of 70\% for $N\geq 8$ when $n_2=500$ and $N\geq 16$ when $n_2=1000$. For ARCH.t and HS the rejection rates are reasonable for all $N$ and more or less in line with the values given by the other two tests; it can be noted that the rejection rates are a little higher when the longer estimation window is used. ARCH.norm is strongly rejected as was the case with the Pearson and Nass tests.

In conclusion, this experiment confirms that using $N=4$ or 8 quantiles gives an effective multinomial test; $N=4$ is appropriate if using a Pearson or Nass tests and $N=8$ gives superior power if using the LRT.

\section{A procedure to implicitly backtest ES} \label{sec:proc-impl-backt} 

 In view of the numerical results obtained in Section~\ref{sec:numerical}, we turn to the question of recommending a test procedure for use in practice.
\begin{itemize}
\item
If we consider simplicity of the test and ease of explaining to management and regulators to be the overriding factors, then a Pearson test with $N=4$ can be recommended. This is a considerable improvement on a binomial test or a test using 2 quantiles, and is easy to implement with standard software. 
\item
The Nass test with $N=4$ or $N=8$ is a less well known test but does appear to be slightly more robust test to the Pearson test in terms of its size properties with comparable power to Pearson for $N=4$ and slightly superior power when $N=8$. Although its power is less than that of the LRT, the difference is relatively small for $N\le 8$ making the Nass a viable and attractive test.
\item
The most powerful test is the LRT which gives good and stable results for $N\geq 4$. However, it requires a sample size for estimation of at least 500 not to be oversized. Moreover, this test requires a little more work to implement as we must carry out an optimization to determine the maximum-likelihood estimates of $\mu$ and $\sigma$ in~\eqref{eq:4}.
\end{itemize}

We can now propose an implicit backtest for ES,  defining a decision criterion based on the multinomial approach we developed so far. Indeed, the ES estimate derived from a model that is not rejected by our multinomial test, is implicitly accepted by our backtest. Hence we can use the same rejection criterion for ES as for the null hypothesis H0 in the multinomial test.

The \citet{bib:basel-16} has proposed a traffic-light system for determining whether capital multipliers should be applied based on the results of a simple exception binomial test based on a backtest length of $n_1=250$ days. We explain how the traffic light system can be extended to any one of our multinomial tests and illustrate the results in the case when $N=2$ (simply because this case lends itself to graphical display).
 
Let $B$ be the number of exceptions at the 99\% level in one trading year of 250 days and let $G_B$ denote the cdf of a Binomial ${\cal B}(250,0.01)$ random variable. In the Basel system, if $G_B(B) < 0.95$, then the traffic light is green and the basic multiplier of 1.5 is applied to a bank's capital; this is the case provided $B \le 4$. If $0.95 \leq G_B(B) < 0.9999$, then the light is yellow and an increased multiplier in the range $[1.70,1.92]$ is applied; this is the case for $B \in \{5,\ldots,9\}$.
If $G_B(B) \geq 0.9999$, the light is red and the maximum multiplier 2 is applied; this occurs if $B \ge 10$. A red light also prompts regulatory intervention.

We can apply exactly the same philosophy. In all our multinomial tests the test statistic $S_N$ (or $\tilde{S}_N$ for the LRT) has an (asymptotic) chi-squared distribution  with some degree of freedom (say $\theta$) under the null hypothesis. Let $G_\theta$ denote the cdf of a chi-squared distribution. If $G_\theta(S_N) < 0.95$, we would set the traffic light to be green; if $G_\theta(S_N) \geq 0.95$, we would set the traffic light to be (at least) orange; if $G_\theta(S_N) \geq 0.9999$, we would set the traffic light to be red. We could easily develop a system of capital multipliers for the orange zone based on a richer set of thresholds.

\begin{figure}
\caption{\label{fig:traffic-lights}Traffic lights based on a trinomial test ($N=2$) with $n=250$, $\alpha_1=0.975$ and
  $\alpha_2 = 0.9875$. $O_1$ and $O_2$ are the numbers of observations falling in the two upper bins (the lower bin contains $O_0 = 250-O_1-O_2$ observations).}
  \begin{center}
\includegraphics[width=10cm,height=10cm]{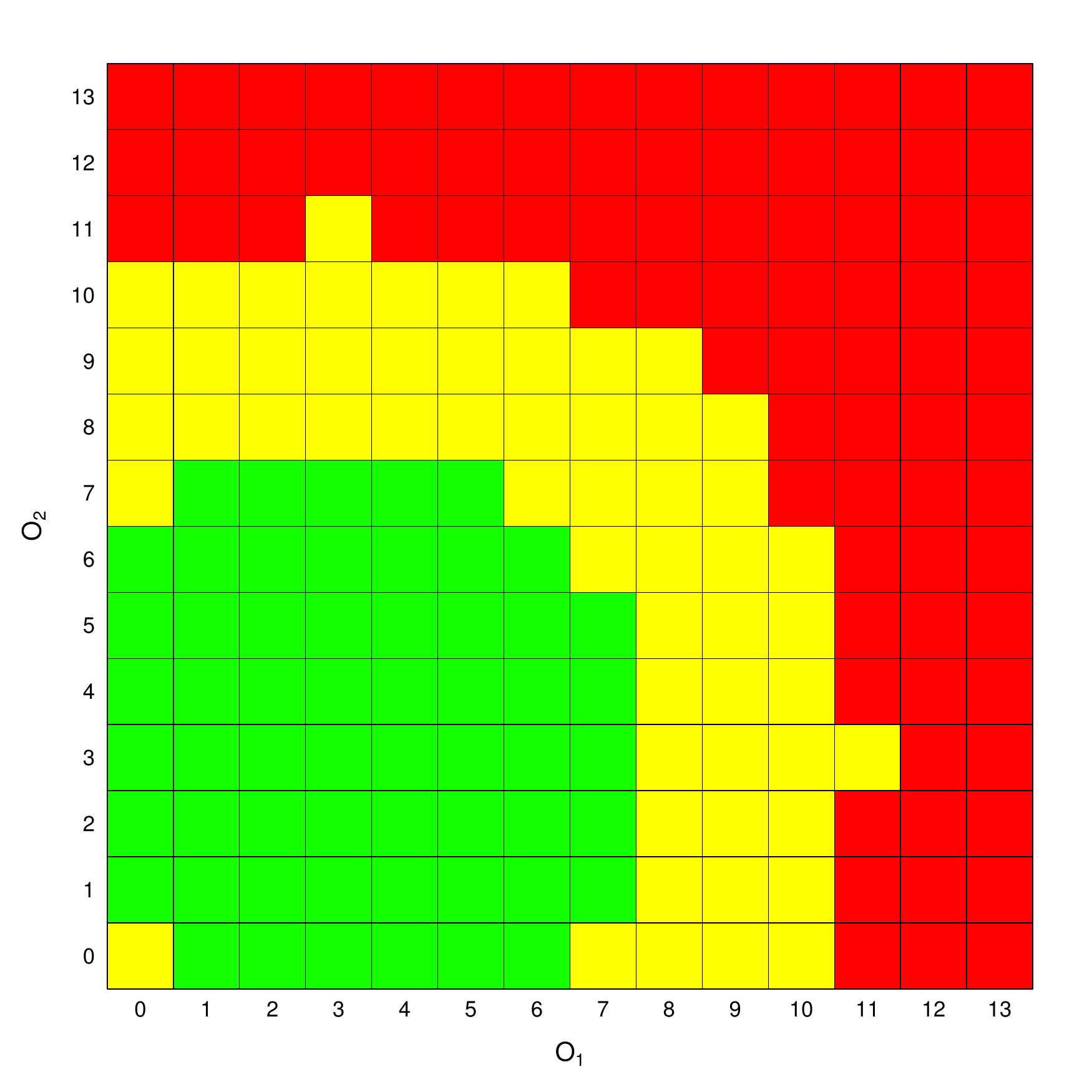}
\end{center}
\end{figure}

 Figure \ref{fig:traffic-lights} shows the traffic-light system for the case when $N=2$ and $n=250$. The particular test that has been used is the Nass test. Obviously values of $N>2$ correspond to cubes and hypercubes that are less easy to display, but we would use the same logic to assign a colour based on the data $(O_0,O_1,\ldots,O_N)$.

\section{Applying our test for model validation on real data} \label{sec:empirical}

To conclude the analysis we apply our multinomial test to real data. 
We consider a hypothetical investment in the Standard \& Poor's 500 index where losses are given by the time series of daily log-returns. We conduct a backtest over the 40-year period from 1976--2016 carrying out a multinomial test in each 4-year period (approximately 1000 days) and comparing the power to that of a one-sided binomial score test of VaR exceptions of the 99\% level; this was the most powerful binomial test considered in Section~\ref{sec:binom-test-results} among tests that had the correct size in samples of size $n=1000$ (see Table~\ref{table:binmultcomparison}). (Note that when we considered a binomial test as a special case of multinomial we considered a two-sided LRT of exceptions at the 97.5\% level, which proved to be a weak test in comparison with multi-level tests.)

For the multinomial test we choose the LRT with $N=8$ equally-spaced levels, starting as usual from $\alpha=0.975$, which was shown to be a particularly powerful test.

In Table~\ref{table:realdata} we give the results for four of the forecasters considered in Section~\ref{sec:dyn}: HS, GARCH.norm, GARCH.t, GARCH.HS. In all methods a rolling window of $n_2=500$ days is used to derive quantile estimates one day out of sample and these are compared with realized losses the next day. The parameters of the GARCH.norm and GARCH.t models are updated every 10 days. This is the same procedure as was used in the dynamic backtesting study of Section~\ref{sec:dyn}. Clearly, in order to initiate the analysis we need 500 days of data pre 1976.

\begin{table}[htbp]
\scriptsize
\centering
\begin{tabular}{llrrlrrrrrrrrrl}
  \toprule
Method & Period & n & B & $p_B$ & $O_{0}$ & $O_{1}$ & $O_{2}$ & $O_{3}$ & $O_{4}$ & $O_{5}$ & $O_{6}$ & $O_{7}$ & $O_{8}$ & $p_M$ \\
  \midrule
HS & 1976-1979 & 1010 &  14 & \cellcolor{green} 0.11 & 988 &   1 &   0 &   1 &   4 &   3 &   5 &   4 &   4 & \cellcolor{green} 0.44 \\
    & 1980-1983 & 1012 &  11 & \cellcolor{green} 0.39 & 983 &   4 &   5 &   6 &   1 &   2 &   1 &   4 &   6 & \cellcolor{green} 0.27 \\
    & 1984-1987 & 1011 &  24 & \cellcolor{red} 0.00 & 969 &   4 &   2 &   1 &   5 &   8 &   8 &   3 &  11 & \cellcolor{red} 0.00 \\
    & 1988-1991 & 1011 &  10 & \cellcolor{green} 0.51 & 991 &   3 &   1 &   1 &   3 &   3 &   1 &   3 &   5 & \cellcolor{green} 0.68 \\
    & 1992-1995 & 1011 &  10 & \cellcolor{green} 0.51 & 991 &   3 &   2 &   2 &   2 &   2 &   5 &   2 &   2 & \cellcolor{green} 0.86 \\
    & 1996-1999 & 1011 &  20 & \cellcolor{yellow} 0.00 & 968 &   4 &   5 &   6 &   5 &   3 &   4 &   6 &  10 & \cellcolor{yellow} 0.01 \\
    & 2000-2003 & 1004 &  14 & \cellcolor{green} 0.10 & 971 &   4 &   4 &   3 &   7 &   1 &   4 &   4 &   6 & \cellcolor{green} 0.28 \\
    & 2004-2007 & 1006 &  17 & \cellcolor{yellow} 0.01 & 977 &   1 &   2 &   3 &   4 &   2 &   3 &   4 &  10 & \cellcolor{yellow} 0.03 \\
    & 2008-2011 & 1009 &  26 & \cellcolor{red} 0.00 & 968 &   5 &   2 &   3 &   3 &   3 &   4 &   8 &  13 & \cellcolor{red} 0.00 \\
    & 2012-2015 & 1006 &   8 & \cellcolor{green} 0.74 & 984 &   2 &   3 &   3 &   3 &   3 &   2 &   2 &   4 & \cellcolor{green} 0.99 \\
    & All & 10091 & 154 & \cellcolor{red} 0.00 & 9790 &  31 &  26 &  29 &  37 &  30 &  37 &  40 &  71 & \cellcolor{red} 0.00 \\ \midrule
  GARCH.norm & 1976-1979 & 1010 &  12 & \cellcolor{green} 0.27 & 981 &   3 &   4 &   2 &   5 &   3 &   3 &   4 &   5 & \cellcolor{green} 0.91 \\
    & 1980-1983 & 1012 &  14 & \cellcolor{green} 0.11 & 981 &   6 &   5 &   0 &   4 &   3 &   5 &   2 &   6 & \cellcolor{green} 0.21 \\
    & 1984-1987 & 1011 &  21 & \cellcolor{yellow} 0.00 & 976 &   2 &   1 &   2 &   6 &   3 &   4 &   5 &  12 & \cellcolor{yellow} 0.00 \\
    & 1988-1991 & 1011 &  17 & \cellcolor{yellow} 0.01 & 983 &   5 &   1 &   2 &   1 &   2 &   3 &   4 &  10 & \cellcolor{yellow} 0.02 \\
    & 1992-1995 & 1011 &  18 & \cellcolor{yellow} 0.01 & 984 &   1 &   3 &   3 &   0 &   2 &   2 &   4 &  12 & \cellcolor{yellow} 0.00 \\
    & 1996-1999 & 1011 &  28 & \cellcolor{red} 0.00 & 968 &   3 &   4 &   1 &   3 &   6 &   3 &   8 &  15 & \cellcolor{red} 0.00 \\
    & 2000-2003 & 1004 &  12 & \cellcolor{green} 0.27 & 975 &   3 &   5 &   2 &   3 &   4 &   4 &   2 &   6 & \cellcolor{green} 0.73 \\
    & 2004-2007 & 1006 &  22 & \cellcolor{red} 0.00 & 967 &   3 &   1 &   3 &   3 &   7 &   4 &   5 &  13 & \cellcolor{red} 0.00 \\
    & 2008-2011 & 1009 &  30 & \cellcolor{red} 0.00 & 959 &   3 &   3 &  10 &   3 &   1 &  12 &   5 &  13 & \cellcolor{red} 0.00 \\
    & 2012-2015 & 1006 &  29 & \cellcolor{red} 0.00 & 963 &   0 &   4 &   3 &   4 &   4 &   6 &   7 &  15 & \cellcolor{red} 0.00 \\
    & All & 10091 & 203 & \cellcolor{red} 0.00 & 9737 &  29 &  31 &  28 &  32 &  35 &  46 &  46 & 107 & \cellcolor{red} 0.00 \\ \midrule
  GARCH.t & 1976-1979 & 1010 &  11 & \cellcolor{green} 0.39 & 981 &   4 &   4 &   1 &   5 &   6 &   1 &   5 &   3 & \cellcolor{green} 0.42 \\
    & 1980-1983 & 1012 &   7 & \cellcolor{green} 0.84 & 985 &   5 &   4 &   2 &   5 &   4 &   1 &   3 &   3 & \cellcolor{green} 0.79 \\
    & 1984-1987 & 1011 &  14 & \cellcolor{green} 0.11 & 977 &   2 &   4 &   5 &   7 &   3 &   3 &   4 &   6 & \cellcolor{green} 0.32 \\
    & 1988-1991 & 1011 &   9 & \cellcolor{green} 0.64 & 984 &   3 &   3 &   6 &   3 &   3 &   0 &   5 &   4 & \cellcolor{green} 0.52 \\
    & 1992-1995 & 1011 &  13 & \cellcolor{green} 0.18 & 985 &   4 &   1 &   1 &   1 &   6 &   5 &   5 &   3 & \cellcolor{green} 0.32 \\
    & 1996-1999 & 1011 &  19 & \cellcolor{yellow} 0.00 & 969 &   6 &   3 &   4 &   4 &   6 &   5 &   7 &   7 & \cellcolor{yellow} 0.05 \\
    & 2000-2003 & 1004 &   8 & \cellcolor{green} 0.74 & 977 &   4 &   2 &   5 &   6 &   3 &   1 &   2 &   4 & \cellcolor{green} 0.58 \\
    & 2004-2007 & 1006 &  20 & \cellcolor{yellow} 0.00 & 971 &   4 &   0 &   4 &   3 &   4 &   9 &   6 &   5 & \cellcolor{yellow} 0.02 \\
    & 2008-2011 & 1009 &  15 & \cellcolor{green} 0.06 & 961 &   4 &  14 &   2 &   9 &   4 &   6 &   5 &   4 & \cellcolor{red} 0.00 \\
    & 2012-2015 & 1006 &  21 & \cellcolor{yellow} 0.00 & 965 &   2 &   5 &   4 &   6 &   5 &   7 &   8 &   4 & \cellcolor{yellow} 0.03 \\
    & All & 10091 & 137 & \cellcolor{yellow} 0.00 & 9755 &  38 &  40 &  34 &  49 &  44 &  38 &  50 &  43 & \cellcolor{red} 0.00 \\ \midrule
  GARCH.HS & 1976-1979 & 1010 &  15 & \cellcolor{green} 0.06 & 979 &   3 &   1 &   5 &   5 &   2 &   0 &   8 &   7 & \cellcolor{yellow} 0.02 \\
    & 1980-1983 & 1012 &   8 & \cellcolor{green} 0.75 & 989 &   4 &   4 &   4 &   1 &   2 &   2 &   2 &   4 & \cellcolor{green} 0.86 \\
    & 1984-1987 & 1011 &  20 & \cellcolor{yellow} 0.00 & 969 &   6 &   1 &   7 &   4 &   4 &   4 &   5 &  11 & \cellcolor{yellow} 0.00 \\
    & 1988-1991 & 1011 &  11 & \cellcolor{green} 0.39 & 986 &   3 &   1 &   5 &   3 &   2 &   3 &   3 &   5 & \cellcolor{green} 0.83 \\
    & 1992-1995 & 1011 &  17 & \cellcolor{yellow} 0.01 & 988 &   0 &   1 &   2 &   2 &   1 &  11 &   4 &   2 & \cellcolor{yellow} 0.00 \\
    & 1996-1999 & 1011 &  14 & \cellcolor{green} 0.11 & 977 &   2 &   5 &   7 &   2 &   4 &   4 &   5 &   5 & \cellcolor{green} 0.32 \\
    & 2000-2003 & 1004 &  14 & \cellcolor{green} 0.10 & 977 &   2 &   4 &   3 &   2 &   3 &   7 &   2 &   4 & \cellcolor{green} 0.58 \\
    & 2004-2007 & 1006 &  21 & \cellcolor{yellow} 0.00 & 972 &   1 &   1 &   8 &   1 &   2 &   4 &   5 &  12 & \cellcolor{red} 0.00 \\
    & 2008-2011 & 1009 &  13 & \cellcolor{green} 0.18 & 981 &   3 &   2 &   4 &   2 &   4 &   2 &   3 &   8 & \cellcolor{green} 0.33 \\
    & 2012-2015 & 1006 &  10 & \cellcolor{green} 0.51 & 978 &   2 &   4 &   6 &   3 &   3 &   3 &   4 &   3 & \cellcolor{green} 0.88 \\
    & All & 10091 & 143 & \cellcolor{red} 0.00 & 9796 &  26 &  24 &  51 &  25 &  27 &  40 &  41 &  61 & \cellcolor{red} 0.00 \\
   \bottomrule
\end{tabular}
\caption{Results of multinomial and binomial backtests
             applied to real data. $B$ gives the number
             of exceedances of the 99\% VaR estimate and the columns $O_0,\ldots,O_N$
             give the observed numbers in cells defined by setting $\alpha = 0.975$.
             $p_B$ and $p_M$ give the $p$-values for a one-sided binomial score
             test and a multinomial LRT respectively.}
\label{table:realdata}
\end{table}

In the table the column marked $B$ gives the number of exceedances of the 99\% VaR estimate and the columns marked $O_0,\ldots,O_N$ give the observed numbers in equally-sized sells above the 97.5\% level; $p_B$ and $p_M$ give the $p$-values for a one-sided binomial score test at the  99\% level and for a multinomial LRT respectively.

We colour the results according to the traffic-lights system described in Section~\ref{sec:proc-impl-backt} both within a single 4-year period and over the whole period. Thus a $p$-value less than 0.05 leads to a yellow colour and a $p$-value less than 0.0001 leads to a red colour; both correspond to rejection of the null hypothesis. It should be emphasized that, in contrast to all the other tables in this paper, Table~\ref{table:realdata} contains $p$-values and not estimates of power or size.

For the HS and GARCH.norm forecasters, the binomial and multinomial tests lead to similar conclusions. The results of the HS forecaster are rejected in 4 out of 10 periods and over the whole period; the results of the GARCH.norm forecaster are rejected in 7 out of 10 periods and over the whole period. The traffic-light colouring is the same for both tests, but $p$-values are generally higher for the multinomial test than for the binomial one.

It is for the other two forecasters that the increased power of the multinomial test is apparent. The multinomial test rejects the GARCH.t forecaster in the period 2008--2011 which contains the 2008 financial crisis, whereas the binomial test does not. The GARCH.t model may be neglecting the asymmetry of the return process in this volatile period and this is uncovered by the multinomial test. The multinomial test also rejects the results of this forecaster over the whole period. The multinomial test also rejects the GARCH.HS forecaster in one additional period (1976--1979) and with higher significance in the period 2004--2007 (red versus yellow).

To conclude, while this is only a single set of results using a single time series, it gives an indication of the increased power of the multinomial test over the binomial. The tabulated numbers between each quantile estimate $O_1,\ldots,O_N$ give additional information about the region of the tail in which the models fail; in a backtest of length $n=1000$ with $N=8$ levels starting at $\alpha =0.975$, the expected numbers are $25/3 \approx 3$ in each cell.

\section{Conclusion}\label{sec:conclusion}

In this  paper we have developed several variants of a multinomial test to simultaneously judge the backtesting performance of trading book models at different VaR levels, and hence to propose an implicit backtest for ES.

We have evaluated the multinomial approach in a series of Monte Carlo simulation studies of size and power and further experiments that replicate typical conditions of an industry backtest. We have also shown the results for an example based on real market risk-factor data.

Carrying out tests with a controlled size is important, particularly for preserving amicable relationships between regulators and banks who would not want their best models incorrectly rejected. However, high power is arguably most important for a regulator who wants a tool for exposing deficient risk models.

As expected, the multinomial test at multiple VaR levels is superior to the standard binomial exception test in distinguishing between good and bad trading book models, particularly in longer backtests. It is able to expose methods that will underestimate tail risk and thus lead to an underestimation of capital according to the ES measure. Our study shows that simultaneously backtesting exceptions at $N=4$ or $N=8$ quantile levels yields a very effective test in terms of balancing simplicity and reasonable size and power properties: it addresses the deficiencies of the binomial test in discriminating between models with different tail shapes and it is clearly superior to a test at two quantile levels, an idea suggested in~\citet{bib:basel-16}.

Our multinomial backtest could easily be performed as a regular routine, in the same way that the binomial backtest is currently carried out based on daily data. We have shown that it is possible to design a traffic-light system for the application of capital multipliers and the imposition of regulatory interventions that is completely analogous to the current traffic-light system based on VaR exceptions over a 250 day period at the 99\% level. We would also suggest moving to longer backtesting periods than 250 days to obtain more powerful discrimination between good and bad backtesting results.

There are of course many other possible backtesting procedures and if developing powerful statistical tests is the main priority, then methods based on realized $p$-values (or PIT values) such as the tests of~\citet{bib:diebold-gunther-tay-98} and~\cite{bib:berkowitz-01} could also be considered. There are also interesting new joint testing procedures of expected shortfall and VaR proposed by~\citet{bib:acerbi-szekely-14} relying on Monte-Carlo simulation.

However, the multinomial tests have the considerable virtue that they are easy to understand, explain and implement as extensions of the standard binomial test. They may thus stand a better chance of gaining acceptance from banks and their regulators.

Finally we note that the multinomial tests may also have a role in extreme value theory (EVT) where they could be considered as a further tool, like the mean excess plot, for distinguishing between light and heavy tails; see~\citet{bib:embrechts-klueppelberg-mikosch-97} for more discussion of statistical methods for analysing extreme data.

\section*{Acknowledgments} 

This project has received the support from the European Union's Seventh Framework Programme for research, technological development and demonstration under grant agreement no 318984 - RARE. 

Marie Kratz acknowledges with gratitude the hospitality of ETH Zurich, RiskLab, as a FIM visitor during Spring 2016, when working on this project. 

Alexander McNeil and Yen Lok acknowledge an insightful dialogue with Michael Gordy about the backtesting of banks' trading models as well as the construction of tests of VaR at multiple levels.

\bibliographystyle{biometrika}

\newcommand{\noopsort}[1]{}

\end{document}